%% file: main.tex
\documentclass[sigplan,screen]{acmart}
\renewcommand\footnotetextcopyrightpermission[1]{}
\usepackage{comment}
\usepackage{enumitem}
\usepackage{framed}
\usepackage{graphicx}
\usepackage{outlines}
\usepackage[linesnumbered,boxed,commentsnumbered]{algorithm2e}
\usepackage{tabularx}
\usepackage{enumitem}

\newenvironment{CompactEnumerate}
  {\begin{enumerate}[leftmargin=*,noitemsep,topsep=0pt]}
  {\end{enumerate}}
\usepackage[style=base,textfont={small, bf},belowskip=-8pt, aboveskip=0pt]{caption}
\input{macros.tex}

\usepackage[small, compact]{titlesec}
\titlespacing{\paragraph}{
  0pt}{
  0.1\baselineskip}{
  1em}

\usepackage[subtle]{savetrees}

\begin{document}

\date{}

\title{\sys{}: Programmed Far-Memory Prefetching for Oblivious Applications}

\newcommand{\UCB}{$^1$}
\newcommand{\NYU}{$^2$}
\newcommand{\ICSI}{$^3$}

\author{Christopher Branner-Augmon\UCB{}, Narek Galstyan\UCB{}, Sam Kumar\UCB{},\\Emmanuel Amaro\UCB{}, Amy Ousterhout\UCB{}, Aurojit Panda\NYU{}, Sylvia Ratnasamy\UCB{}, Scott Shenker\UCB{}\ICSI{}}
\affiliation{
    \institution{\UCB{}UC Berkeley, \NYU{}NYU, \ICSI{}ICSI}
    \country{}
}

\renewcommand{\shortauthors}{Branner-Augmon, Galstyan, Kumar, Amaro, Ousterhout, Panda, Ratnasamy, and Shenker}
\acmConference[]{\eat}{\eat}{}

\begin{abstract}
Using memory located on remote machines, or far memory, as a swap space is a promising approach to meet the increasing memory demands of modern datacenter applications. Operating systems have long relied on prefetchers to mask the increased latency of fetching pages from swap space to main memory. Unfortunately, with traditional prefetching heuristics, performance still degrades when applications use far memory. In this paper we propose a new prefetching technique for far-memory applications. We focus our efforts on memory-intensive, oblivious applications whose memory access patterns are independent of their inputs, such as matrix multiplication. For this class of applications we observe that we can perfectly prefetch pages without relying on heuristics. However, prefetching perfectly without requiring significant application modifications is challenging.

In this paper we describe the design and implementation of \textbf{\sys{}}, a system that provides \textbf{p}re-\textbf{p}lanned \textbf{p}refetching for general \textbf{o}blivious applications.
We demonstrate that \sys{} can accelerate applications, e.g., running them 30-150\% faster than with Linux's prefetcher with 20\% local memory.
We also use \sys{} to understand the fundamental software overheads of prefetching in a paging-based system, and the minimum performance penalty that they impose when we run applications under constrained local memory.
\end{abstract}

\maketitle
\vspace{-0.14in}
\input{01_introduction.tex}
\input{02_background.tex}
\input{03_overview.tex}
\input{04_design.tex}
\input{05_implementation.tex}
\input{06_evaluation.tex}
\input{07_related.tex}
\input{08_conclusion.tex}

\section*{Acknowledgments}

Emmanuel Amaro was partially supported by a UCMEXUS-Conacyt Fellowship.
This work was funded by NSF Grants 1817115, 1817116, 1704941, and was supported by Intel, VMware, and Microsoft.
This material is based upon work supported by the National Science Foundation Graduate Research Fellowship Program under Grant No. DGE-1752814.
Any opinions, findings, and conclusions or recommendations expressed in this material are those of the authors and do not necessarily reflect the views of the National Science Foundation.

\bibliographystyle{plain}
\bibliography{reference}

\end{document}

%% file: macros.tex
\newcommand{\sys}{3PO}

\definecolor{darkgreen}{rgb}{0.3, 0.9, 0}

\newcommand{\allnotes}[1]{}
\renewcommand{\allnotes}[1]{#1}

\newcommand{\eat}[1]{}

\newcounter{graphcounter}

\newcommand{\secref}[1]{\S{}\ref{#1}}
\newcommand{\figref}[1]{Figure~\ref{#1}}

\newcommand{\para}[1]{\smallskip\noindent{\bf #1}}
\newcommand{\us}{$\mu$s}

%% file: 01_introduction.tex
\section{Introduction}\label{sec:introduction}

The rise of memory-intensive big-data applications such as machine learning~\cite{spark} has caused memory demands in datacenters to increase drastically~\cite{datacenter_as_a_computer}.
At the same time, the slowing of Moore's Law means that memory costs (per GB) are no longer decreasing~\cite{kang2014co, lee2016technology}.
This has two consequences: first, applications often want to access more memory than is available on their local server and second, datacenter operators are incentivized to use their available memory as efficiently as possible.
Recently \emph{far memory} has emerged as a solution to these challenges, enabling applications to make use of unused memory elsewhere in a cluster~\cite{cfm, aifm, leap, infiniswap}.
By moving some data to far memory when local memory is full, far memory systems allow applications to run with less local memory and improve overall resource utilization.

Unfortunately, while existing far-memory systems~\cite{leap, cfm} enable good performance at high local memory ratios,\footnote{The {\em local memory ratio} is the fraction of an application’s total memory that is allowed to reside in local memory.} their performance degrades significantly as you decrease the amount of local memory. For example, with Fastswap, application runtime degrades by 11-226\% with 40\% local memory~\cite{cfm}.
A fundamental reason for this degradation is that far memory has higher latency than local memory (a few microseconds vs. about 100~ns).
One way to reduce overhead is to more carefully decide what data to move from local memory to far memory when local memory is full, in order to reduce the number of memory accesses that stall on far memory.
However, \emph{prefetching} is a more promising approach because it has the potential to prevent \emph{all} stalls on far memory.
Unfortunately, state-of-the-art prefetching algorithms~\cite{hashemi2018learning, lynx, learningrelaxedbelady, leap, he2013io, soundararajan2008context, diskseen} rely on heuristics, which are imperfect and often behave suboptimally.
Therefore, they are not effective enough to fully mask the latency of far memory.

However, we observe that some applications---including linear algebra operations (e.g., matrix multiply), certain machine learning algorithms, and Fourier transforms---have a special property that admits heuristic-free, nearly optimal prefetching.
This special property is that \emph{the sequence of memory accesses issued by such programs is independent of the program's inputs}.
For deterministic applications of this type, we can determine the memory access pattern up front and then use it to guide prefetching when the program is run on any input.

This special property is called {\em obliviousness} and it has been studied primarily in the context of computer security~\cite{oram, traceoblivious, obliviousds, opaque}.
Obliviousness is useful for security because it guarantees that no information about a program's input is leaked via memory-related side channels.
One recent prior work, MAGE~\cite{mage}, goes further and exploits this obliviousness for memory management, including prefetching.
Though MAGE is limited to a family of cryptographic applications called Secure Computation, we observe that the implications of obliviousness for memory management extend beyond these security-related settings.
In this context, we ask the question: \textbf{can we enable nearly perfect prefetching of far memory for \emph{general} oblivious applications, with little to no modification?}
To answer this question, we design and implement \sys{}, a system that provides \textbf{p}re-\textbf{p}lanned \textbf{p}refetching for \textbf{o}blivious applications.
\sys{} builds on the operating system's virtual memory subsystem, treating far memory as swap space.
Using \sys{}, we (1) explore how much the oblivious applications mentioned above can benefit from programmed prefetching, and (2) push prefetching to the limit and study the fundamental software overheads of prefetching in a paging-based virtual memory subsystem.

In designing \sys{}, we could not directly leverage designs from prior work.
MAGE's techniques and design are specialized to Secure Computation; it assumes that the program is written in a DSL and that it is acceptable to run it in a software interpreter.
Other approaches to application-directed prefetching assume that the program is built with a particular compiler~\cite{mowry1996automatic} or that the programmer provides hints to the prefetcher~\cite{vandebogart2009reducing, patterson1995informed, tomkins1997informed}.
These approaches are not suitable for \sys{}, which aims to support general oblivious applications with little to no modification.
Instead, \sys{} works in two phases, as follows.
First, \sys{} observes a program as it executes (with a given input) and generates a {\em tape} indicating which pages of memory need to be fetched as this program executes.
Second, when the program is run later on a different input,
\sys{} prefetches pages from far memory, using the tape as a guide.
Realizing this design required overcoming two main challenges.

\sys{}'s first challenge is generating a program's tape.
To do this, \sys{} introduces a \emph{tracer}, which records the pages accessed by a program as it executes.
Existing tools based on dynamic binary instrumentation (DBI)~\cite{pin, valgrind} can collect such a trace but are not efficient; they are known to take impractically long and result in a trace that is impractically large~\cite{mage}.

We address this challenge with two insights.
The first insight is that, for the purpose of page-based prefetching, page-granularity traces (rather than access-granularity) are sufficient.
This enables us to design an efficient in-kernel tracer that ensures that it incurs a page fault each time a new page is accessed and then records these page accesses in the page fault handler.
We improve on this further with our second insight: to organize the trace not as a series of page accesses, but as a series of \emph{microsets}---small working sets.
With microsets, our tracer records sets of pages that are accessed consecutively, but does not record the complete sequence of accesses within each set of pages.
This accelerates tracing and produces smaller traces.
Microsets omit information about the exact sequence of page accesses within a microset, yet we find that we can use microsets that are up to hundreds of pages in size without degrading prefetching accuracy.
\sys{} additionally \emph{post-processes} each trace to filter out pages that are likely to already be in local memory, leaving a more concise \emph{tape} of pages to be prefetched at runtime. Note that the tape only specifies which pages to prefetch; \sys{} lets Linux's default eviction policy determine which pages to evict.

\sys{}'s second challenge is to ensure that prefetching is timely.
\sys{} does not have visibility into where the program is in its execution, so it is difficult for \sys{} to know when it should prefetch the next page on the tape.
Prior approaches circumvent this by having the application software manage memory (e.g., in a software interpreter~\cite{mage}) or by relying on a compiler~\cite{mowry1996automatic} or developer~\cite{app_directed, informed_prefetching, informed_multi_process} to insert prefetch directives.
But such techniques are unsuitable for supporting generic applications with minimal modifications, as \sys{} aims to do.

To solve this problem, \sys{} identifies pages in the tape that are guaranteed to page fault when accessed (because they are not present), and selects a subset of them as \emph{key pages}.
Key pages serve as synchronization points from which to initiate prefetching, so \sys{} aims to space them evenly along the tape.
Specifically, \sys{} uses the page fault on each key page as a signal to select the next key page and to prefetch all pages that will be accessed in between.
\sys{} generalizes both its tracing and prefetching techniques to support multithreaded applications, with a separate tape and separate key pages for each thread.

We implemented a prototype of \sys{} in the Linux 4.11.0 kernel, and evaluated it by combining the \sys{} prefetcher with Fastswap~\cite{cfm}, a state-of-the-art far-memory system.
We found that \sys{} can collect the memory-access trace of a program orders of magnitude faster than Intel Pin~\cite{pin}, a state-of-the-art tool for dynamic binary instrumentation.
Using \sys{}, we can run applications at much lower local memory ratios than with Linux or state-of-the-art prefetchers such as Leap~\cite{leap}, or run them at the same local memory ratio with better performance. For example, with \sys{}, applications run 30-150\% faster at 20\% local memory than with Linux's prefetcher.

Despite \sys{}'s benefits, we find that we cannot completely prevent performance degradation at low local memory ratios.
Overheads such as minor page faults on prefetched pages and TLB flushes for multithreaded applications still cause runtimes to degrade, for example by 0-300\% at a local memory ratio of 20\%.
Even so, we believe that \sys{} can be a useful tool for oblivious applications and that it sheds light on the fundamental limitations of how much performance we can hope to salvage with better prefetching.

%% file: 02_background.tex
\section{Background}\label{sec:background}
\vspace{-.12in}
\subsection{Far Memory}

Far memory is an emerging datacenter design that allows operators to improve
cluster resource utilization by leveraging the unused memory of remote servers
or memory blades. In a typical setup, an oversubscribed machine fetches memory
over a low-latency network using RDMA, ~\cite{liang2005swapping,infiniswap,cfm}
or
even traditional TCP~\cite{aifm}. Applications can leverage far memory using
custom APIs~\cite{aifm, remote_regions}
or in ways that remain transparent to applications~\cite{cfm,
infiniswap, leap, legoos}; this transparency is achieved by using the existing virtual memory subsystem to swap pages.
\sys{} takes the
latter approach because of its goal of generality. Indeed, some cloud operators argue that adopting specialized APIs for far memory in large scale clusters is impractical~\cite{softwaredefinedfarmem}.

Running applications that swap at low local memory ratios can impose very high bandwidth requirements.
In the worst case, if the allowed local memory for an application is close to zero, then the network bandwidth would need to be comparable to the local memory bus bandwidth---e.g., 800 Gbps on one modern platform~\cite{memorybandwidth}---in order to achieve performance similar to that of only using local memory.
Thus, running applications at low local memory ratios would be infeasible with older storage technologies such as HDDs, even with perfect prefetching hiding access latency, because they provide less than 5 Gbps~\cite{fastesthdd}. In contrast, recent high-bandwidth non-volatile memory technologies can achieve tens of Gbps per memory module~\cite{optaneperf}, while network bandwidth can exceed 100 Gbps.
These technologies make it feasible to explore running applications with severely restricted local memory. However, as an application's local memory ratio decreases, accurate and timely prefetching become more important.

\subsection{Memory Management in Linux}

Most modern processors manage memory at the granularity of pages, typically 4 KB each. When a program tries to access a page, the processor checks its {\em page table entry} (PTE) to see if it is marked as ``present'' in local memory, and the CPU trigger a {\em page fault} if the page is ``not present''. The operating system is responsible for handling faults by fetching the faulted page from a swap device. When fetching pages from swap, Linux will also prefetch a dynamically-sized batch of pages that are consecutive (either in swap space or virtual address space) to the page that triggered the fault.

Linux's paging mechanisms impose overheads on applications primarily in two ways. First, if an application tries to access a page that Linux failed to prefetch, it incurs a {\em major page fault} and must block until the page has been fetched from the swap device and mapped into the process's address space. Second, if a page is present but not yet mapped into a process's address space (e.g., because it was prefetched but has not yet been accessed), the first access to it will incur a {\em minor page fault}. Handling minor page faults entails some software overhead but typically does not require waiting on I/O. Most prefetching algorithms aim to eliminate overheads from major page faults, though as we will discuss (\S\ref{ss:prefetching}), oblivious applications provide an opportunity to reduce the overheads of minor page faults as well.

\subsection{Application Requirements} \label{ss:obliviousness}

\sys{} requires two main properties from applications: (1) {\em obliviousness}, meaning that application's memory access patterns are independent of their inputs, and (2) {\em deterministic} memory access patterns at page granularity. Obliviousness originates from security contexts, where it is useful because it can guarantee that a program does not leak information via memory-related side channels~\cite{oram, traceoblivious, obliviousds, opaque}. Instead, \sys{} leverages obliviousness for its potential to improve prefetching performance. As a result, \sys{}'s requirements for applications differ slightly from those for traditional oblivious applications. First, because \sys{} fetches memory at the granularity of pages, it requires only page-level obliviousness rather than access-level obliviousness; an application that might reorder accesses within a page is sufficiently oblivious for \sys{}. Second, \sys{} can eliminate some inputs that might prevent an application from being traditionally oblivious (e.g., matrix size) by retracing the application for each of these inputs (e.g., generate one tape for each target matrix size). 

We observe that many memory-intensive applications today meet these requirements. Examples include linear algebra operations such as matrix multiplication, determinant computation, and eigenvalue computation. In addition, Fourier transforms, and some graphics workloads (e.g., ray tracing, marching cubes) and machine learning algorithms (e.g., neural networks) also have these properties. \sys{} targets these workloads.

%% file: 03_overview.tex
\section{\sys{}'s Design}

\sys{}'s goal is to enable nearly perfect prefetching of far memory for oblivious applications, without requiring significant application modifications. To limit modifications to the application, we build \sys{} into the Linux kernel and leverage Linux's machinery for page-based memory management, rather than requiring applications to use a new API as in some existing approaches~\cite{aifm, mage, remote_regions}. By ``nearly perfect'' we mean that memory accesses should almost never stall because they are waiting for memory to be swapped in, even when running applications with very limited local memory. Ideally \sys{} would also achieve this with low overhead, allowing an application to achieve the exact same runtime when run with a low local memory ratio (e.g., 5\%) as when run with 100\% local memory. While \sys{} is able to provide high accuracy, we find that it does entail some overheads which prevent it from achieving performance that is completely independent of the local memory ratio~(\S\ref{ss:overheads}).

\begin{figure}[t]
    \centering
    \includegraphics[width=\linewidth]{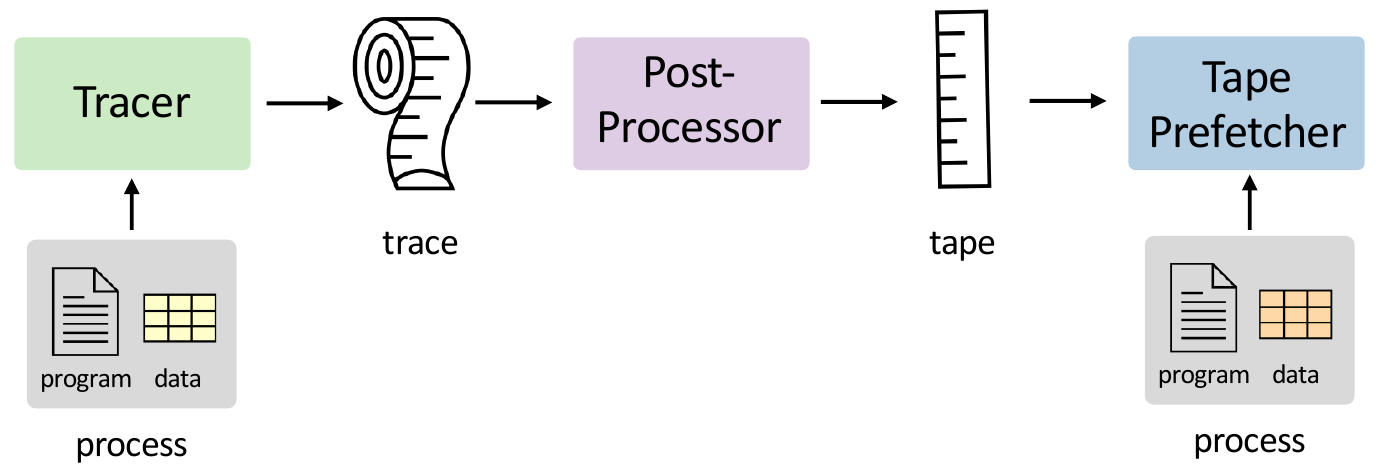}
    \caption{\sys{} (1) runs an oblivious program with input data, using its {\em tracer} to collect a memory-access {\em trace}, (2) {\em post-processes} the trace to create a concise {\em tape}, and (3) runs the same program with different input data, prefetching pages according to the tape.}
    \label{fig:overview}
\end{figure}

There are two main challenges in designing \sys{}: (1) obtaining a {\em tape} for an application describing which pages to prefetch, and (2) prefetching pages at the correct times during execution, i.e., ensuring that prefetching is timely. \sys{} addresses these challenges with a three-step process, as shown in \figref{fig:overview}. First, before actually running an oblivious program, a user executes it offline with sample input data,
using \sys{}'s in-kernel {\em tracer} to record the program's sequence of memory accesses, or {\em trace}. When the user later executes their program online, many accesses in the tape will not trigger prefetches, because the pages will be resident in memory from earlier accesses in that run. Thus \sys{}'s second step is for the user to feed the trace through \sys{}'s post-processor, which simulates which pages are evicted  given a specific target local memory ratio, in order to generate a more concise {\em tape} of pages to prefetch during execution. This reduces overheads during prefetching, because the prefetcher does not have to scan through a long list of pages that are already present. Finally, the user can run their program with different input data under constrained local memory, with \sys{}'s prefetcher loading the tape and accurately prefetching according to it.
Note that \sys{} assumes that the amount of available local memory remains fixed throughout the execution of the program.

\sys{} uses Linux's default eviction policy, which is similar to LRU.
One could extend \sys{}'s current design to use Belady's optimal MIN~\cite{beladys} algorithm to precompute what pages should be evicted; we leave this to future work.
Next we describe each step of \sys{}---tracing (\S\ref{ss:trace}), post-processing (\S\ref{ss:tape}), and prefetching (\S\ref{ss:prefetching})---in more detail. We describe how \sys{} handles single-core applications, and then explain how we extend the techniques to multicore in~\S\ref{ss:multi_core}.

%% file: 04_design.tex
\subsection{Collecting a Memory Trace}\label{ss:trace}
When collecting a trace of a program's memory accesses, \sys{} aims to: (1) collect a concise trace, (2) collect it quickly, and (3) support general programs---programs that are not written in a specialized way for memory-trace collection.
Existing general-purpose tracing options based on dynamic binary instrumentation (DBI)~\cite{pin, valgrind, transparent_dynamic} have very high overheads and are slow enough that it is impractical to trace even small programs this way. 
This is because DBI-based approaches involve instrumenting most memory accesses
which adds significant overhead, particularly for memory-intensive applications that can access memory very frequently.
\sys{} aims to retain the generality of DBI but at higher efficiency, by allowing most memory accesses to execute without software intervention.
We achieve this benefit via two techniques we discuss next: page-granularity tracing and microsets.

Note that \sys{}'s tracing techniques apply to regular applications as well as those that are data oblivious. Thus we believe that \sys{}'s tracer is useful independent of \sys{}, as a tool for helping researchers and developers better understand the memory-access patterns of their programs.

\subsubsection{Page-Granularity Tracing}\label{s:page-granularity}

Because operating systems typically fetch memory (from storage or far memory) in units of pages, it is sufficient to record traces at page-granularity.
By \emph{page-granularity}, we mean that consecutive accesses to the same page are condensed into a single entry in the trace.
This approach not only makes the trace smaller, but also allows us to leverage the processor's hardware page table support to only perform tracing work when an application accesses a different page, rather than on most memory accesses.

\sys{}'s high-level approach to page-granularity tracing is to modify the page-fault handler to record all page accesses, and to force a program to incur a page fault each time it accesses a different page. This approach incurs many more page faults than a program would normally incur during execution, but allows \sys{} to record a complete trace of page accesses.

More precisely, the tracer maintains an invariant over $S$, the set of pages whose accesses are recorded in a trace. The invariant is that \emph{of all pages in $S$, only the most recently accessed page has its ``present'' bit set in its page table entry (PTE)}.
If the next access to a page in $S$ is to a different page than the most recently accessed page, the memory management unit will observe that the page is not present and raise a page fault.
This allows \sys{}'s tracer to record the access in the trace, mark this page as ``present,'' and mark the previous page as ``not present''.
Alternatively, if the next access to a page in $S$ is to the most recently accessed page, then the access proceeds without a page fault, thereby coalescing all consecutive accesses into a single entry in the trace with no additional tracer overhead.

This approach poses a challenge: how can \sys{} distinguish between pages that are actually present but have been marked as ``not present'' by the tracer, and those that are truly not present?
For example, when a process accesses a page for the first time, the Linux page fault handler needs to execute to allocate a page frame to back the page; simply setting the ``present'' bit in the PTE would cause incorrect behavior.
To distinguish between these two cases, when \sys{}'s tracer marks the PTE of a present page as ``not present'', it also sets a special bit in the PTE, which we call the \sys{} bit. On a page fault, the tracer can then check if the \sys{} bit is set and use this to determine if it can simply set the ``present'' bit to true and return, or if this page must be handled by Linux's regular page-fault handler. 

The invariant of at-most-one-page-present-at-once poses challenges with instructions that can access multiple pages at once, such as \texttt{movdqu} or \texttt{vscatterqpd}.
Though we relax this invariant below (\S\ref{ss:microsets}), these instructions require that a specific set of pages are present simultaneously, which requires special care.
For example, \texttt{movdqu} can require that up to two pages are present simultaneously.
\sys{} handles this case by recognizing when a sequence of page faults on the same instruction alternates between two pages (e.g., pages ABAB) and responds by temporarily marking both pages as ``present'' until the instruction pointer advances.
This technique handles instructions that access two pages at once. It can be extended to support instructions that can access more than two pages, such as \texttt{vscatterqpd}; we leave this to future work.

The tracer provides a system-call-based interface for processes to specify when tracing begins and ends (\S\ref{s:impl}); pages allocated outside of this interval will be excluded from $S$. Processes should invoke the syscall to start tracing as soon as possible during startup and before allocating the large memory buffers they will use for computation (e.g., on the heap), so that accesses to that memory will be included in the trace. This approach has tradeoffs: it may miss some page accesses that need to be prefetched during execution, but it also allows tracing to skip memory accesses that occur only during process initialization. In addition, to speed up tracing and focus only on the large memory buffers used for computation, the tracer explicitly excludes stack pages from $S$ by checking which virtual memory area an access falls in. The tracer also ignores page faults due to instruction fetches by checking a bit in the page-fault error code.

\para{Assumptions.}
The current implementation of \sys{}'s tracer assumes that during tracing, page faults only occur for two reasons: (1) self-induced page faults caused by marking present pages as ``not present'' or (2) page faults due to memory allocation. It may be possible for the tracer to handle other special minor page faults, such as those due to copy-on-write resulting from a call to \texttt{mmap}. This could be done by detecting a minor page fault on a page that has its \sys{} bit set and is already marked as present, and then executing the regular Linux page-fault handler; we leave this to future work.

\sys{} could similarly support tracing in the presence of major page faults, i.e., when a process's memory does not fit in available local memory.
While a page is swapped out of memory, Linux repurposes its PTE to record information about where the page is stored, so the \sys{} bit may be overwritten and cannot be used to determine if a page is physically present or not. However, \sys{} could leverage other mechanisms to detect if a page was swapped out, and still rely on the \sys{} bit for pages that remained physically present.

\IncMargin{0.5em}
\begin{algorithm}[t]
  \footnotesize
  \SetAlgoLined\SetAlgoNoEnd
  \SetFuncSty{}
  \SetStartEndCondition{ }{}{}
  \SetKwProg{Fn}{def}{\string:}{}
  \SetKwFunction{Range}{range}
  \SetKw{KwTo}{in}\SetKwFor{For}{for}{\string:}{}
  \SetKwIF{If}{ElseIf}{Else}{if}{:}{else if}{else:}{}
  \SetKwFor{While}{while}{:}{fintq}
  \SetKwFunction{GetRequestStartTime}{GetRequestStartTime}
  \SetKwFunction{Sleep}{sleep}
  \SetKw{Continue}{continue}
  \SetKw{Return}{return}
  \setlength{\AlCapSkip}{1em}

  $microset = \{\}$ \\
  \Fn{on\_page\_fault(page p)}{
    \tcp{record access to p}
    \If{size of microset == MICROSET\_SIZE} { \label{alg:full_start}
        \tcp{start a new microset}
        \For{p' in microset}{
            append $p'$ to trace \\
            clear present bit for $p'$ \\
        }
        $microset = \{\}$ \\  \label{alg:full_end}
    }
    add $p$ to $microset$ \\ \label{alg:add}
    
    \quad{}\\
    \tcp{resolve page fault}
    \If{p's \sys{} bit is set} {
        \tcp{skip normal page-fault handling}
        set $p$'s present bit \\ \label{alg:present}
    }
    \Else {
        \tcp{first access to p}
        set $p$'s \sys{} bit \\
        run normal page-fault handling \\
    }
    \Return{}
  }
  \caption{The main logic in \sys{}'s tracer.
  }\label{alg:tracing_microsets}
\end{algorithm}\DecMargin{0.5em}

\begin{figure}[t]
    \centering
    \includegraphics[width=\linewidth]{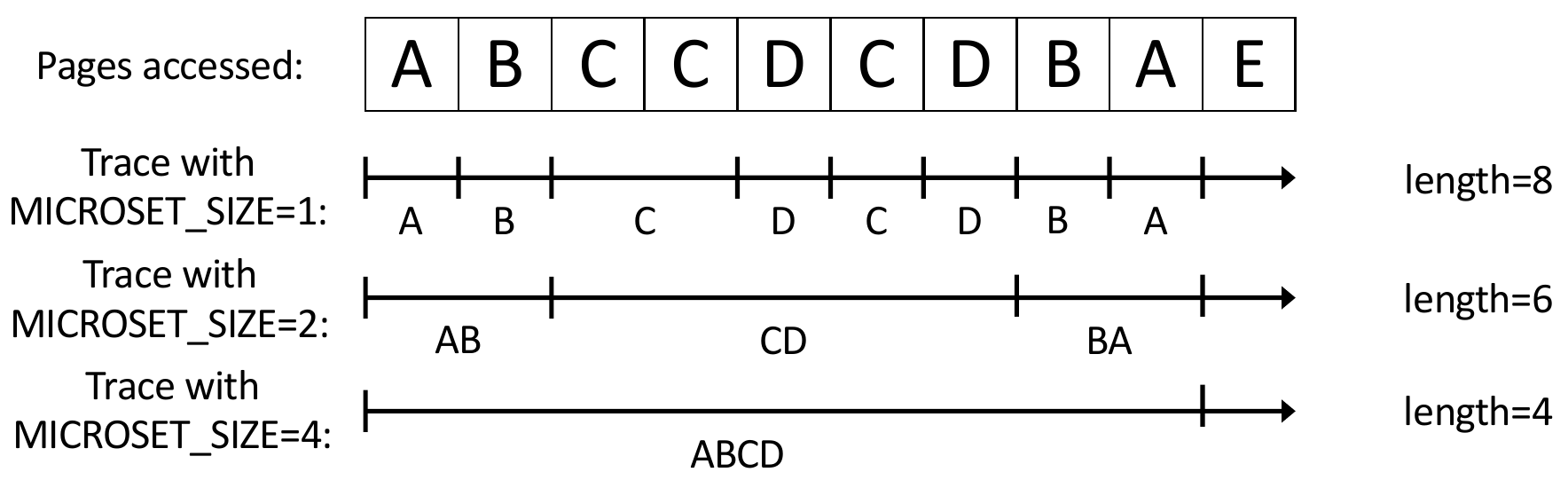}
    \caption{A sequence of accessed pages and resulting traces with different microsets sizes.\vspace{-.15in}}
    \label{fig:microsets}
\end{figure}

\subsubsection{Microsets}\label{ss:microsets}

While page-granularity tracing provides a significant improvement relative to a DBI-based solution, it still wastes time and space by recording information at significantly higher granularity than necessary for prefetching.
In practice, when a program runs, more than one page will be resident in local memory at once, and tracing accesses for these already-present pages provides no useful information for the prefetcher.
For example, if a program accesses the sequence of pages ABABAB, as long as the amount of local memory is at least two pages, A and B should remain present, so the last four accesses will not trigger prefetches and are therefore not useful to record.

Thus to improve efficiency, \sys{}'s tracer relaxes the invariant described above (\S\ref{s:page-granularity}) to allow multiple pages---a small {\em microset} of up to \texttt{MICROSET\_SIZE} pages---to remain marked as ``present'' simultaneously. As shown in Algorithm~\ref{alg:tracing_microsets}, the microset starts out empty and each time a new page is accessed, triggering a page fault, the tracer adds it to the microset (line~\ref{alg:add}) and marks it as present (line~\ref{alg:present}). When a new page is accessed and the current microset is full (lines~\ref{alg:full_start}-\ref{alg:full_end}), the tracer adds all pages in the microset to the trace, marks them as ``not present'', and resets the microset to empty. Then when pages in that microset are accessed again, they will incur minor page faults and be added to the microset again.

With microsets, each trace consists of a sequence of microsets where each microset represents a small working set of \texttt{MICROSET\_SIZE} pages that are accessed consecutively, as illustrated in Figure~\ref{fig:microsets}. During tracing, only \texttt{MICROSET\_SIZE} page faults are incurred per microset, and the remaining accesses proceed without software intervention. In most cases increasing the microset size yields a shorter trace, and tracing time decreases commensurately with the length of the trace.

How large should \texttt{MICROSET\_SIZE} be? At one extreme, \texttt{MICROSET\_SIZE} could be the number of pages in the smallest local memory size that we intend to use for this application. However, microsets this large degrade prefetching accuracy. The reason for this is that \sys{} assumes that at runtime, Linux will not evict any page in a microset between when the page is first and last accessed as part of that microset; if a page was evicted during this interval, it would not be prefetched before its next access and would incur a major page fault. In practice this assumption holds as long as \texttt{MICROSET\_SIZE} is not too large, so that pages within a microset are accessed close together in time. However, because Linux's eviction policy is not strictly LRU, this assumption may not hold with very large values of \texttt{MICROSET\_SIZE}. Overall we found that a wide range of microset sizes are able to strike a good balance, significantly accelerating tracing without degrading accuracy (\S\ref{ss:tape_eval}); by default \sys{} uses a \texttt{MICROSET\_SIZE} of 1024 pages.

\subsection{Computing the Tape from the Trace}\label{ss:tape}

In principle, \sys{} could accurately prefetch data at runtime using the memory trace produced in~\S\ref{ss:trace} directly, by interpreting it as a list of pages to prefetch. However, this would result in needlessly attempting to prefetch many pages that are already present in local memory from an earlier access. For example, in Figure~\ref{fig:microsets} with a microset size of 2, if our local memory contained at least 4 pages, \sys{} probably would not need to prefetch BA again, because they would still be in memory from the earlier accesses to AB. To avoid prefetching BA again, \sys{} must traverse the page table to discover whether the pages are already local, adding overhead. In addition, despite the techniques in~\S\ref{ss:trace}, the redundant accesses in the trace can make it large enough (e.g., gigabytes) that traversing it at runtime would consume significant memory and CPU cycles.\footnote{One could in theory prefetch the trace, but this would save memory at the expense of network bandwidth and would still consume CPU cycles.}

Therefore, we perform a {\em post-processing} step on the trace to filter out pages that are likely to be in local memory and therefore do not need to be prefetched.
Conceptually, this transforms the trace, which is a sequence of accessed pages structured as microsets, into a \emph{tape} describing what to prefetch at runtime.
The tape allows \sys{} to avoid extraneous prefetching and reduces the amount of memory the prefetcher must traverse at runtime~(\S\ref{ss:tape_eval}).

The set of pages that remains in memory at runtime depends on three factors: the prefetching algorithm, the eviction policy, and the amount of available local memory. Thus to generate a tape, \sys{}'s post-processor traverses the trace, simulating \sys{}'s perfect prefetching algorithm and an eviction policy with a particular target local memory size to determine which pages will not be present and will need to be prefetched. Ideally \sys{} would simulate Linux's eviction policy, since this is what will be used at runtime. Unfortunately, Linux's eviction policy is quite complex and depends on the timing of different events~\cite{linux_kernel}, so it would be difficult to simulate it accurately. Instead, \sys{} simulates a simple LRU eviction policy. Linux's eviction policy bears some resemblance to LRU but differs in many ways; despite this we found that simulating LRU instead is accurate enough to avoid major page faults in practice~(\S\ref{ss:tape_eval}).

A user should run post processing to generate a separate tape for different local memory ratios they would like to be able to run their program at. However, we have found that tapes generated for a specific target local memory ratio can be used for executions at slightly larger local memory ratios; this adds a small runtime overhead from scanning through the extra pages on the tape, but does not degrade accuracy~(\S\ref{ss:tape_eval}). For example, a user could generate tapes at increments of 10\% local memory, and round their local memory ratio down when choosing which tape to use.

\vspace{-.11in}
\subsection{Prefetching using the Tape}\label{ss:prefetching}

The goal of the \sys{} prefetcher is to use the tape to bring pages from remote memory to local memory ahead of time, thereby avoiding major page faults. Though the tape provides information about exactly which pages need to be prefetched, it has no information about when to prefetch them. For example, a na\"ive prefetcher could prefetch all of an application's remote memory at once according to the tape as soon as the application started, but this would provide no benefit since the prefetched pages would be evicted to make room for subsequent prefetches, before any of them were accessed. Thus the main challenge faced by the prefetcher is to provide {\em timely} prefetching. This involves two components. First, the prefetcher needs to stay {\em synchronized} with the application, that is, it must know roughly where along the tape the application currently is in its execution. Second, the prefetcher must fetch each batch of pages {\em in advance} so that the pages are likely to arrive by the time the application accesses them.

\para{Synchronization.} In order to provide synchronization, \sys{} introduces {\em key pages}. At any given time, the prefetcher maintains one key page for each application. When choosing a key page, \sys{} chooses a page that will be accessed in the near future and that will trigger a page fault (because it is not currently mapped).
When a page fault occurs on a key page, the prefetcher can resynchronize, updating its state about where the application is in its execution along the trace. In addition, the prefetcher will prefetch the next batch of pages and choose the next key page at this time. Though the prefetcher prefetches key pages, it does not mark them as present, so they will still trigger a minor page fault when accessed.

\sys{} chooses each key page by scanning forward along the tape to find a page that is guaranteed to trigger a page fault. Because the prefetcher fetches pages in batches, it starts looking for the next key page at the index of the current key page plus \texttt{BATCH\_SIZE}. However, it is possible that this page is already currently present and mapped, and will thus not trigger a page fault. This can occur because the post processing uses an approximate rather than an exact model of Linux's eviction policy, or due to other threads executing concurrently (\secref{ss:multi_core}). In either case, the prefetcher scans forward along the tape after the current key page plus \texttt{BATCH\_SIZE}, checking if each page is mapped by consulting the page table, until it finds a page that is unmapped and uses this as the next key page.\footnote{In an alternate approach, you could always advance the key page by \texttt{BATCH\_SIZE} and force it to trigger a page fault by marking its PTE as ``not present'', but this would add overheads from TLB shootdowns.}

\para{Prefetching in advance.} When a page fault occurs on a key page, the prefetcher needs to decide which pages to prefetch next. One approach to do this would be to start with the next page on the tape after the key page and prefetch all pages through the next key page. However, this approach does not prefetch pages far enough in advance and causes delayed hits~\cite{delayed_hits}. These can occur due to either the latency overhead of fetching pages or due to bandwidth limitations. For latency, each page prefetch takes several microseconds to complete, so if the program accesses the first page after the key page in the meantime, it will suffer a delayed hit and have to block until the page arrives. The bandwidth needed for prefetching pages can also be bursty over time~(\S\ref{ss:requirements}), so that even if the first prefetched page in a batch arrives on time, later pages may not if the NIC is not fast enough (we observed this phenomenon often with 10 Gbits/s NICs).

\begin{figure}[t]
    \centering
    \includegraphics[width=1.0\linewidth]{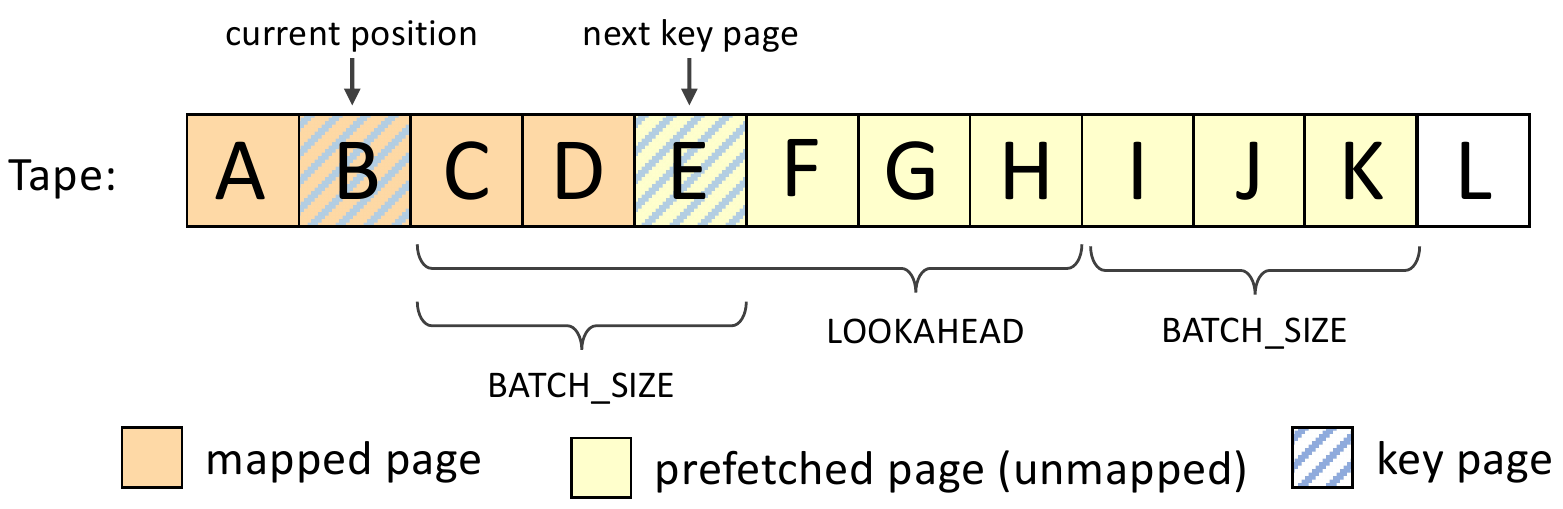}
    \caption{The state of pages after the prefetcher finishes handling a fault on key page B. It chooses E as the next key page, maps all pages before E, and prefetches pages through K.}
    \label{fig:prefetching}
\end{figure}

To reduce delayed hits, \sys{} prefetches pages in advance of when they are needed. As Figure~\ref{fig:prefetching} shows, the \texttt{LOOKAHEAD} parameter determines how far in advance pages are prefetched. When \sys{} needs to prefetch the next batch of pages, it will begin prefetching from the first page in the tape it has not yet prefetched, and fetch all pages up to
\texttt{LOOKAHEAD} + \texttt{BATCH\_SIZE} pages after the key page it just faulted on.

\para{Reducing prefetching overheads.}
\sys{}'s focus on oblivious applications provides a unique opportunity to reduce the overheads of prefetching. When Linux prefetches pages, it brings them into local memory but does not map them into the process's address space; this way Linux will not charge an incorrectly prefetched page to a process's memory limit. Additionally, the prefetched page may not even belong to the current process. The first access to a prefetched page will then trigger a minor page fault, the process will switch into the kernel, and the kernel will map the page into the process's address space.
In contrast, \sys{} can assume that prefetched pages belong to the current process and will always be accessed, so it can map them in advance of their actual access: when handling a page fault on a key page, \sys{} maps the pages from this key page up until the page before the next key page, as shown in Figure~\ref{fig:prefetching}.
This avoids extra mode switches in and out of the kernel and reduces the overhead of prefetching.

\subsection{Multi-Threaded Applications} \label{ss:multi_core}

An oblivious single-threaded application may no longer be oblivious when parallelized, if the set of memory accesses issued by each thread is not deterministic.
For example, if work is partitioned dynamically between threads at runtime (e.g., with work stealing~\cite{intel_tbb, go, java_fork_join}), then the resulting program is not oblivious.
However, in a restricted class of multi-threaded applications---those that partition work statically and deterministically across threads---each thread can remain oblivious.
Libraries for parallel computation such as OpenMP~\cite{openmp} and IntelTBB~\cite{intel_tbb} support parallelizing computation in this way.
\sys{} aims to support these statically-partitioned multi-threaded oblivious applications.

Even when work is statically partitioned across threads, perfect prefetching is challenging because threads may run in a non-deterministic order (e.g., thread A runs first during tracing but thread B runs first at runtime). This can cause the complete sequence of page accesses (across all threads) to also be non-deterministic. \sys{} addresses this by handling each thread separately. \sys{} collects a separate tape for each thread, post-processes each tape individually to generate per-thread tapes, and prefetches for each thread using its own tape.

While this is conceptually straightforward, additional challenges arise because multiple threads may access the same page.
During tracing, if thread A accesses a page while it is already mapped due to an access by thread B, it will not trigger a page fault and that page will be omitted from A's trace. This can cause a major page fault at runtime if thread A happens to access that page before thread B.
\sys{} addresses this by pinning all threads to the same core during tracing, so that multiple threads cannot concurrently access the same page.
Counter-intuitively, this approach also makes tracing faster. This is because the tracer modifies PTEs frequently (\S\ref{ss:trace}), and each modification requires a TLB shootdown to invalidate any cached copies of this PTE on other cores. These shootdowns add significant overhead~\cite{optimizing_tlb} and we found that it is much faster to trace programs on a single core, where we can leverage cheaper core-local TLB invalidations instead.

During post processing, the target amount of local memory is a key input parameter, but \sys{} does not know what fraction of the target local memory will be utilized by each of the $N$ threads. \sys{} addresses this by post processing each tape with one $N$\textsuperscript{th} of the target local memory.
The intuition is that each thread is entitled to roughly one $N$\textsuperscript{th} of the memory if threads operate on disjoint memory and perhaps more if they share memory; we have found that this approach works well in practice~(\S\ref{ss:tape_eval}).

Finally, during prefetching at runtime, concurrent threads risk breaking \sys{}'s synchronization mechanism, by violating the guarantee that each thread's key page will cause that thread to incur a page fault.
For example, suppose thread A chooses its next key page, but before thread A accesses it, thread B accesses that page.
B will incur a page fault and will map A's key page in the process' virtual address space, preventing thread A from incurring a page fault when it accesses the page.
Thus, the prefetcher will lose track of A's position on its tape.
\sys{} takes special measures to prevent this from happening.
Each time a thread B maps a page, it checks if this page is a key page for another thread A, and if it is, B advances A's key page to the next unmapped page on B's tape. To prevent a race condition in which one thread maps a page concurrently with another thread selecting that page as its key page, threads use a lock-protected variable to declare that they are about to map a page before doing so, so that other threads can identify the page as ineligible for use as a key page.

%% file: 05_implementation.tex
\section{Implementation} \label{s:impl}

\sys{}'s implementation consists of two parts. First, a kernel component comprised of 1400 lines of C code added to the Linux kernel memory subsystem; this code exposes a system call API to applications (Table~\ref{t:sys_interface}), creates and manages trace files within the Linux kernel, and prefetches from remote memory according to a tape. Second, a trace post-processing tool written in 250 lines of Python. To use \sys{}, a user first uses the \texttt{sys\_begin(RECORD)} syscall to generate a trace for their application. Next, they run the post processor to generate a tape. Finally, they use the \texttt{sys\_begin(PREFETCH)} syscall to load the tape and run the program with \sys{}'s prefetcher. Users can also use the \texttt{sys\_begin(AUTO)} syscall to avoid recompiling their program between tracing and execution.

As we mentioned in \S\ref{s:page-granularity}, our tracer sets a \sys{} bit and marks as ``not present'' the PTEs of the pages accessed by the traced application. Using the special bit this way is done carefully because if the kernel sees our special bit when it does not expect it, it could cause unexpected behavior or sofware bugs. Therefore, we also modified the kernel to make sure that when tracing, the \sys{} bit is cleared before any kernel procedure reads a PTE, and is set again afterwards.

We considered leveraging multiple different existing far-memory systems for far-memory support; each has pros and cons. One system, Fastswap~\cite{cfm}, offloads the work of choosing pages to evict and writing them to far memory to a separate ``reclaimer core'', accelerating applications by freeing application cores from this task in most cases. However, Fastswap performs evictions synchronously, only allowing one outstanding write per core and wasting CPU cycles on evictions. As a result, applications are throttled because a single reclaimer core cannot handle all evictions for an application when evicting one page at a time, even at moderate local memory ratios.  Another far-memory system, Leap~\cite{leap}, overcomes this by supporting asynchronous evictions, but does not offload reclamation to a separate core. To achieve the best of both systems, we built on top of Fastswap, but augmented it with support for asynchronous evictions, changing about 80 lines of code. This allows us to handle evictions for several applications with a single reclaimer core (\S\ref{ss:requirements}).

For simplicity, \sys{}'s implementation does not support transparent huge pages or address space layout randomization; we disable both when tracing and running applications.

We have tested our \sys{} implementation with C++ binaries and with Python programs that use \texttt{numpy} (see Table~\ref{t:applications}). For the Python applications, we disable the garbage collector by calling \texttt{gc.disable()} at the beginning of the application to avoid its non-deterministic memory accesses. We experimented with another Python library, \texttt{scikit-learn}, but found that it yielded traces that are not oblivious, even for oblivious algorithms. One possible cause is that Python import statements trigger calls to \texttt{mmap}, which can map memory in different locations in different runs.

\begin{table}
\footnotesize
\begin{center}
\begin{tabular}{|c|p{5cm}|}
\hline
Function & Description\\
\hline
\texttt{sys\_begin(RECORD)} & Begin recording a trace for the current process.\\
\hline
\texttt{sys\_begin(PREFETCH)} & Begin prefetching with an existing tape.\\
\hline
\texttt{sys\_begin(AUTO)} & If the kernel does not have a trace for this binary, generate it. Otherwise use existing tape to prefetch.\\
\hline
\texttt{sys\_end()} & Indicates end of trace recording or prefetching. Resources get freed here.\\
\hline
\end{tabular}
\end{center}
\vspace{.1in}
\caption{\sys{}'s tracing and prefetching system-calls.}
\vspace{-.2in}
\label{t:sys_interface}
\end{table}

%% file: 06_evaluation.tex
\section{Evaluation}\label{sec:evaluation}

\begin{table*}[t]
\begin{center}
\footnotesize
 \begin{tabularx}{2\columnwidth}{X|X|p{1.6cm}|p{11cm}}
 Workload & Language & Max RSS (MB) & Description  \\ \hline \hline
  dot\_prod & C++ & 2000 & computes the dot product of two vectors using Eigen~\cite{eigen} \\ 
  mvmul & C++ & 2100 & multiplies a square matrix by a vector using Eigen~\cite{eigen} \\
  matmul & C++ & 400 & multiplies two matrices using Eigen~\cite{eigen}   \\
  matmul\_$p$ & C++ & 400 & same as matmul, but parallelized over $p$ threads using static work partitioning with OpenMP~\cite{openmp} \\
  sparse\_mul & C++ & 1200 & multiplies two square matrices in a sparse representation using Eigen~\cite{eigen}, matrices are 90\% zeroes \\
  np\_matmul & Python & 400 & multiplies two square matrices using the \textsf{numpy} library~\cite{numpy} \\
  np\_fft & Python & 4100 & computes a discrete Fourier transform using the \textsf{numpy} library~\cite{numpy}
 \end{tabularx}
\end{center}
\vspace{0.2in}
\caption{Applications used for evaluation.
Resident set sizes (RSS) are rounded to the nearest 100 MB.}
\label{t:applications}
\vspace{-1em}
\end{table*}

In evaluating \sys{}, we focus on four main questions:
\begin{CompactEnumerate}
\item How do applications perform with \sys{}, compared to with other prefetchers? (\S\ref{ss:runtime})
\item How do network characteristics impact prefetching performance with \sys{}? (\secref{ss:network})
\item What prevents \sys{} from achieving constant performance as we decrease the local memory ratio? (\S\ref{ss:overheads})
\item What are \sys{}'s CPU and network requirements? (\S\ref{ss:requirements})
\item How fast are \sys{}'s tracer and post processor, and how are speed and accuracy impacted by parameters such as \texttt{MICROSET\_SIZE}? (\S\ref{ss:tape_eval})
\end{CompactEnumerate}

\para{Applications.} We evaluate \sys{} using the seven applications listed in Table~\ref{t:applications}. For each application, we measure its maximum memory usage (or resident set size) when running with unlimited local memory (using \texttt{/usr/bin/time -v}) and use this as the amount of memory it is allocated at a local memory ratio of 100\%. The only modification we applied to each application was to invoke the \sys{} syscalls.

\para{Systems evaluated.} We evaluate four systems:
\begin{CompactEnumerate}
\item \textit{\sys{}}, our proposed system. Unless stated otherwise, we use a \texttt{MICROSET\_SIZE} of 1024 pages, \texttt{BATCH\_SIZE} of 100 pages, and \texttt{LOOKAHEAD} of 400 pages. We found that these values performed well across different local memory ratios and across the applications we evaluated.
\item \textit{Leap}~\cite{leap}, a state-of-the-art far-memory prefetcher.
\item \textit{Linux (Fastswap*)}, which consists of Linux's default prefetching policy and our version of Fastswap, augmented to support asynchronous evictions.
\item \textit{Linux (Leap)}, which consists of Linux's default prefetching policy using Leap's Infiniswap-based RDMA backend~\cite{infiniswap} for fetching pages from far memory.
\end{CompactEnumerate}
For all systems we use the most recent Linux kernel version they support; for \textit{\sys{}} and \textit{Linux (Fastswap*)} we use version 4.11, and for \textit{Leap} and \textit{Linux (Leap)} we use version 4.4.125. Kernel versions $<$4.14 prefetch a window of pages that are contiguous to the faulted page in \textit{swap space}. In versions $\geq$4.14 there is second prefetching policy that builds the window of pages to prefetch based on each application's \textit{virtual address space}. We did not evaluate the latter as none of the systems in our evaluation support it. 

\para{Experimental setup.}
We ran experiments on CloudLab. To evaluate how different network latencies and bandwidths impact performance, we used three different setups.
\begin{enumerate}[leftmargin=*,topsep=0in,noitemsep]
\item \textit{25gb}. For our default setup we use xl170 machines with Intel E5-2640v4 CPUs, connected with a 25 Gbps network. Reading a 4 KB page over this network takes 5.0 \us{}.
\item \textit{10gb}. While most prior far-memory research has focused on settings within a single rack, we also evaluate what happens if far memory is placed in a different rack, leading to higher latencies to access it. For these experiments, we use programmable Dell-S4048 switches to build networks where the xl170 machines are separated by multiple switches; these switches only support 10 Gbps bandwidth. Reading a 4 KB page takes from 5.5 \us{} with 0 intervening switches (\textit{10gb\_0switch}) to 15.2 \us{} with 4 intervening switches (\textit{10gb\_4switch}).
\item \textit{56gb}. Finally, we were unable to use the Leap kernel with the xl170 instances, so we evaluate Leap with c6220 instances instead. These instances have 2 Intel Xeon E5-2650v2 CPUs and are connected by a 56 Gbps Infiniband fabric where reading one page takes 3.4 \us{}.
\end{enumerate}

We use Linux's \textsf{cgroup} feature to limit the physical memory available to each application to induce memory pressure.

\subsection{Application Performance}\label{ss:runtime}

\begin{figure}[t]
    \centering
    \includegraphics[width=\linewidth]{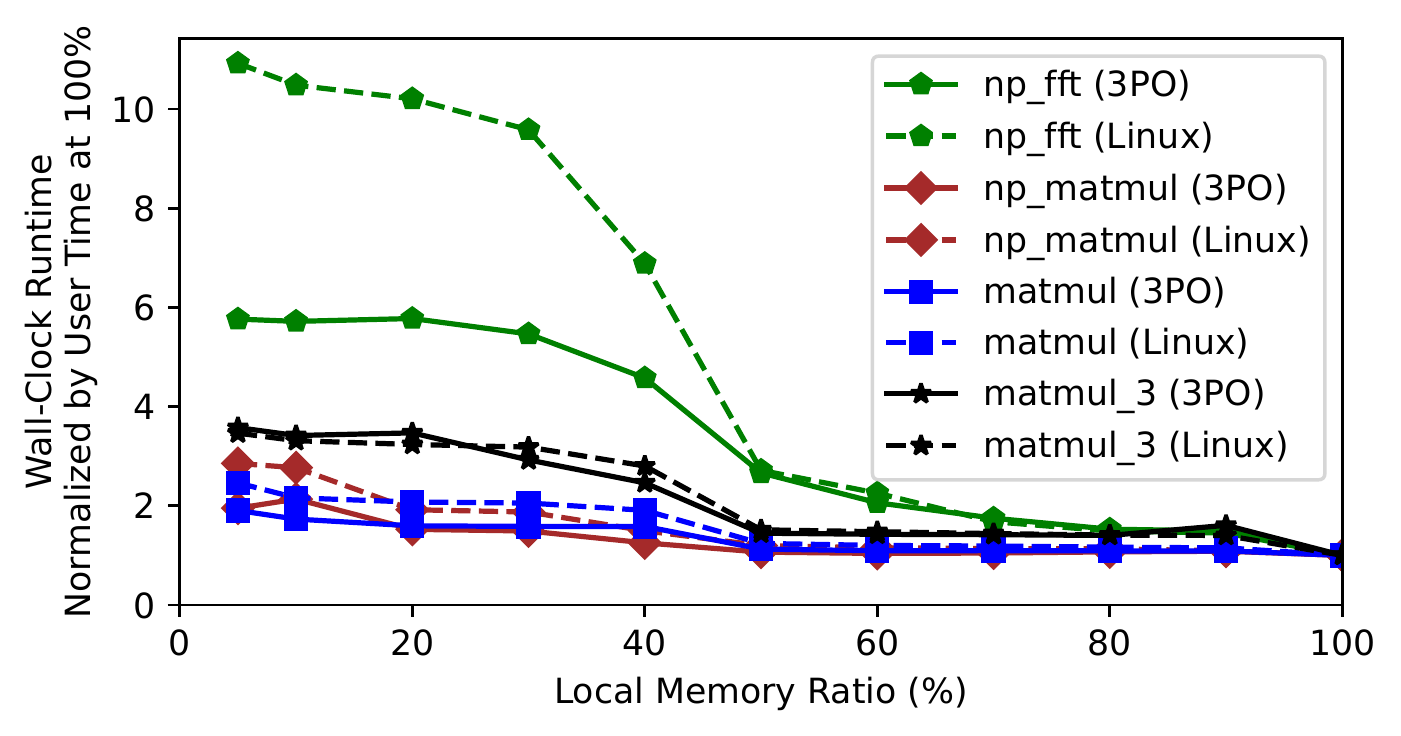}
    \caption{Runtime vs. local memory ratio (see also \figref{fig:sweep_local_memory_2}).}
    \label{fig:sweep_local_memory_1}
\end{figure}

\begin{figure}[t]
    \centering
    \includegraphics[width=\linewidth]{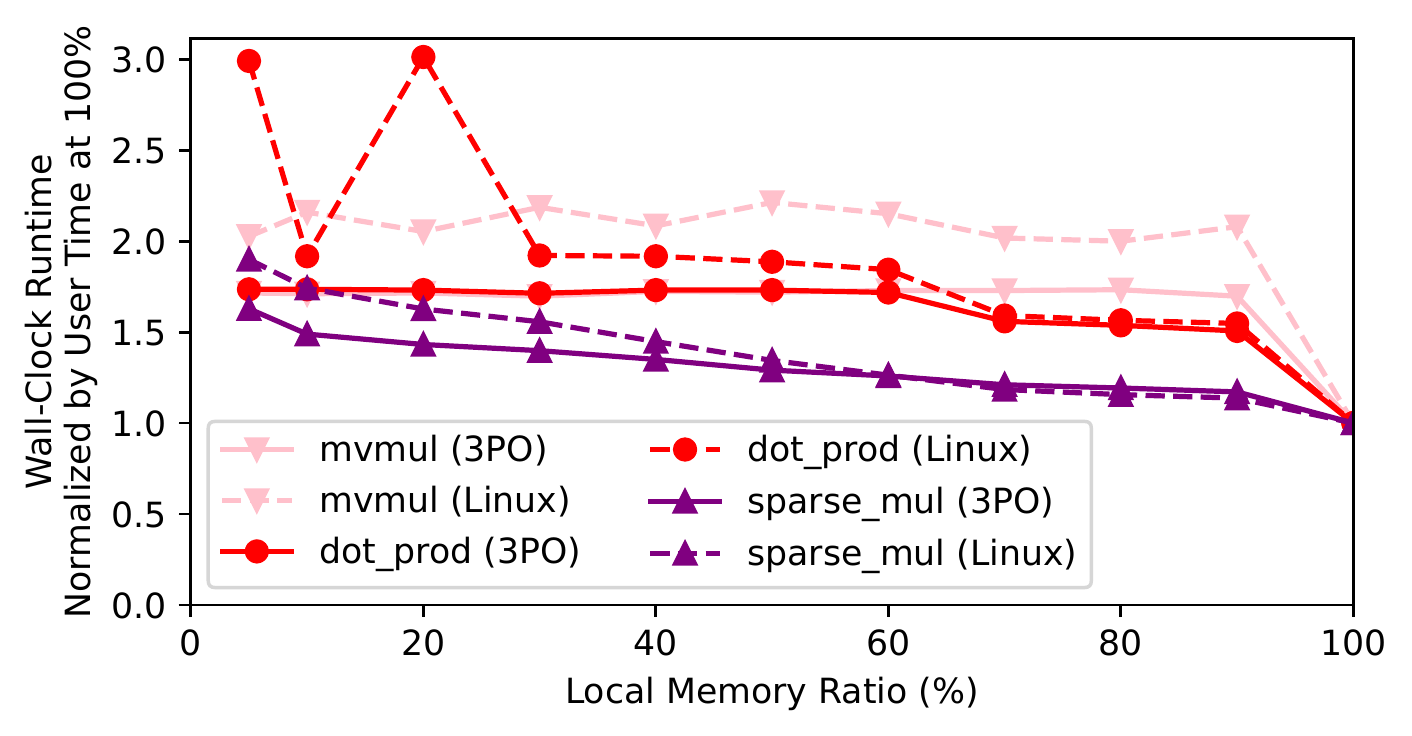}
    \caption{Runtime vs. local memory ratio (see also \figref{fig:sweep_local_memory_1}).}
    \label{fig:sweep_local_memory_2}
\end{figure}

\begin{figure}[t]
    \centering
    \includegraphics[width=\linewidth]{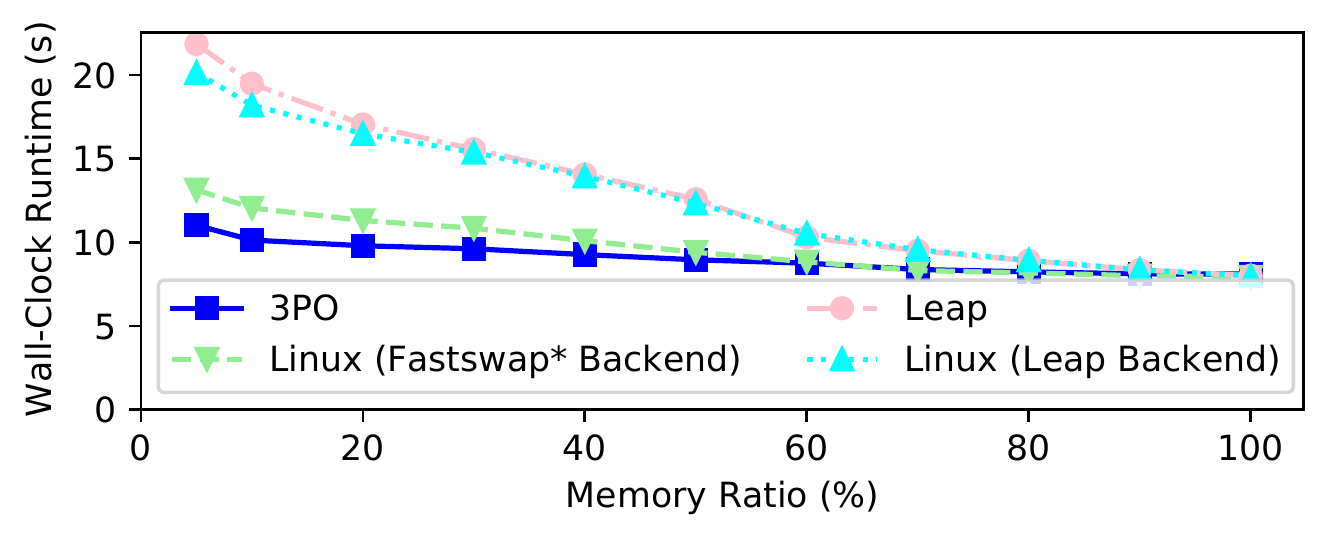}
    \caption{Wall-clock runtimes for sparse matrix multiply.}
    \label{fig:leap_wallclock}
\end{figure}

We begin by comparing the performance of \sys{} and Linux (Fastswap*) for a variety of workloads and local memory ratios (Figures~\ref{fig:sweep_local_memory_1} and \ref{fig:sweep_local_memory_2}).\footnote{\sys{} slightly increases the memory footprint of programs, meaning that the wall-clock time at 100\% local memory may exceed the user time at 100\% local memory. Therefore, we normalize the runtime by the \emph{user time} at 100\% local memory, rather than the wall-clock runtime at 100\% local memory. The only exception is that, at 100\% local memory, we always report the ratio as 1, to indicate ``no degradation.''}
Overall, \sys{} provides a measurable improvement over Linux on most workloads.
\sys{}'s speedup varies depending on the application and memory ratio, with the speedup generally higher at lower memory ratios.
This is because memory accesses go to far memory more frequently at low memory ratios, making performance more sensitive to the prefetching algorithm that is used.

Next, we evaluate how \sys{}'s performance compares to Leap~\cite{leap}, a state-of-the-art prefetching system.
\figref{fig:leap_wallclock} shows the wall-clock runtime of the matrix multiply workload for all 4 setups, as we vary the amount of local memory from 100\% (no memory pressure) to 5\% (95\% percent of resident set in far memory). Due to the previously mentioned Leap limitations, we use c6220 instances connected by a 56 Gbps network for this evaluation.
In this case we find that Leap and Linux (Leap) perform similarly, and \sys{} and Linux (Fastswap*) perform similarly, suggesting that prefetching algorithms have a minimal impact on runtime.
This is to be expected due to the low-latency network used for this evaluation.
The performance difference between Linux (Leap) and Linux (Fastswap*) is comparatively larger due to offloaded evictions in Fastswap*, which suggests that the choice of RDMA backend is more important than the prefetching policy for this hardware setup.
Although \figref{fig:leap_wallclock} only shows the results for \textit{sparse\_mul}, these same observations hold for other workloads as well.

\begin{figure}[t]
    \centering
    \includegraphics[width=\linewidth]{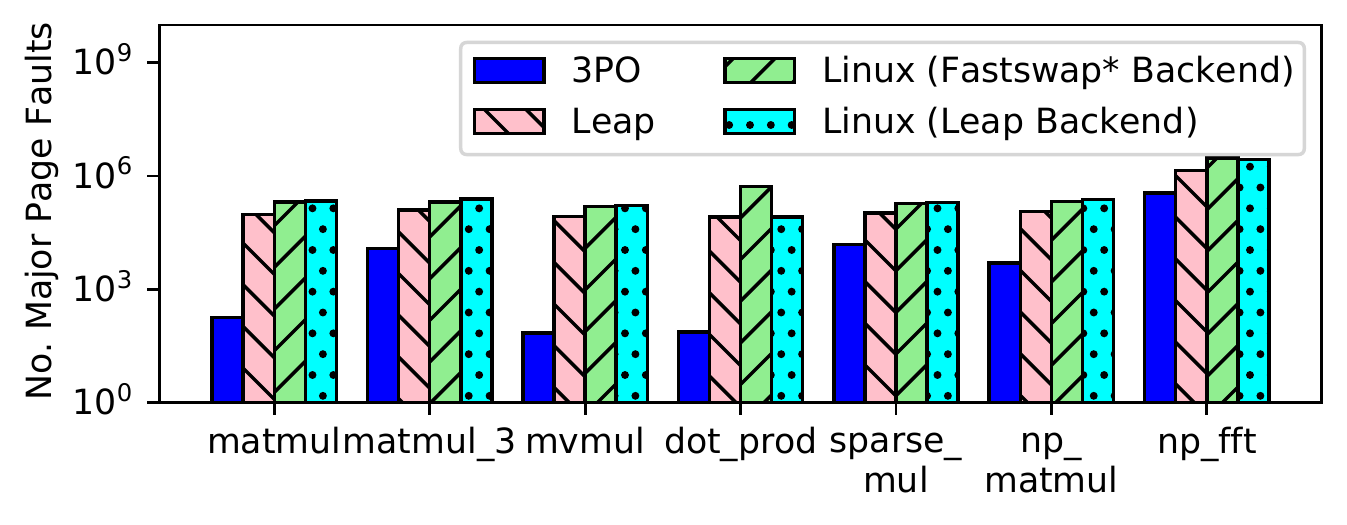}
    \caption{Major page fault counts at a 30\% memory ratio}
    \label{fig:leap_pagefault_by_workload}
\end{figure}

To understand the impact of prefetching in a way that is less dependent on the hardware setup, particularly the latency of RDMA, we measure the number of major page faultsperating at a 20\% memory ratio.
\figref{fig:leap_pagefault_by_workload} compares, at a 30\% memory ratio, the number of major page faults experienced by Leap and \sys{} for various workloads.\footnote{Linux major page fault numbers are sometimes bimodal. For example, in \figref{fig:leap_pagefault_by_workload} the number of major page faults for Linux (Fastswap* Backend) for \emph{dot\_prod} were often similar to those for Linux (Leap Backend).}
Note the log-scale vertical axis.
\sys{}'s major page fault count is consistently orders of magnitude smaller than Leap's.
This suggests that \sys{} may significantly outperform Leap on higher-latency networks where major page faults impose higher overheads.

\subsection{Impact of Network Characteristics} \label{ss:network}
In a datacenter deployment, far memory may be located in a different rack than the machine accessing it.
This would result in higher latency to access far memory, impacting the performance implications of \sys{}'s prefetching.
To explore this, we measure the speedup of \sys{} and Linux in the \textit{25gb}, \textit{10gb\_0switch}, and \textit{10gb\_4switch} setups.

\begin{figure}[t]
    \centering
    \includegraphics[width=\linewidth]{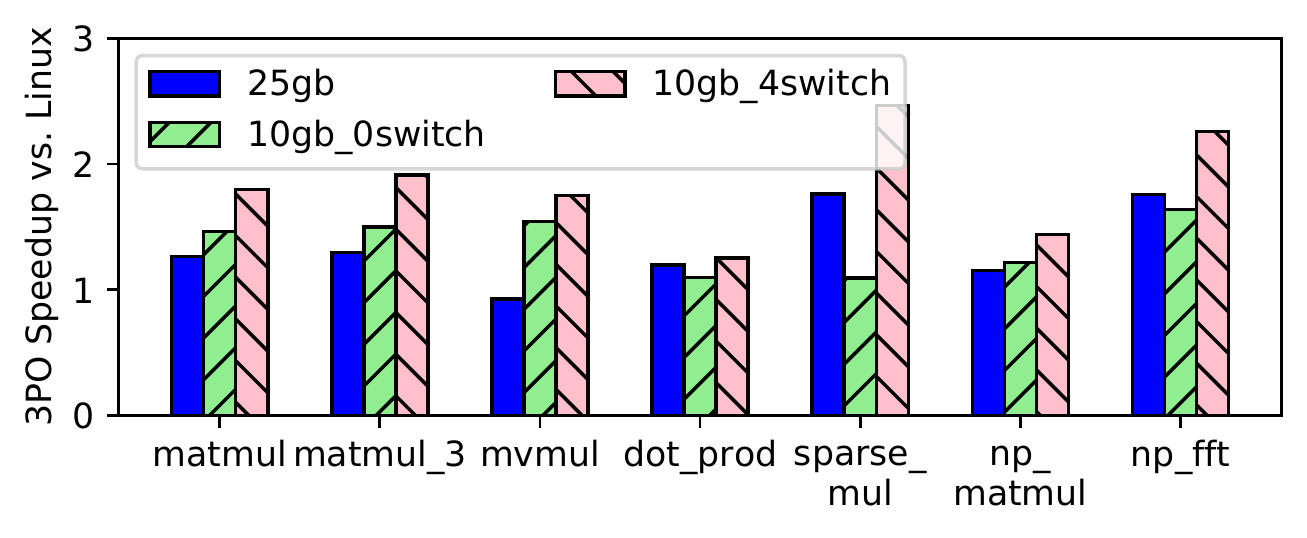}
    \caption{\sys{}'s speedup over Linux for various workloads and network configurations, at a 20\% memory ratio}
    \label{fig:sweep_network}
\end{figure}

The results, shown in \figref{fig:sweep_network}, show that \sys{} has larger speedups on higher latency networks.
This is because the cost of a major page fault is higher on such networks, causing perfect prefetching to have a bigger impact on performance.
For some workloads, such as \emph{dot\_prod}, \sys{} has a higher speedup on \emph{10gb\_0switch} than on \emph{25gb}, even though \emph{25gb} has lower latency.
The reason is that the network bandwidth becomes a bottleneck.
Once the network is saturated, the latency of accessing far memory increases due to queuing delays, causing performance to degrade due to delayed hits.
This affects both \sys{} and Linux, causing the speedup of \sys{} relative to Linux to decrease for the \emph{10gb} setup for these workloads.

Overall, at a 20\% memory ratio, \sys{}'s speedup compared to Linux ranges from $0.9\times$ to $1.8\times$ on the low-latency \textit{25gb} setup.
On the higher-latency \textit{10gb\_4switch} setup, the speedup increases to $1.3\times$ to $2.5\times$.

\subsection{\sys{} Overheads} \label{ss:overheads}

\begin{figure}[t]
    \centering
    \includegraphics[width=\linewidth]{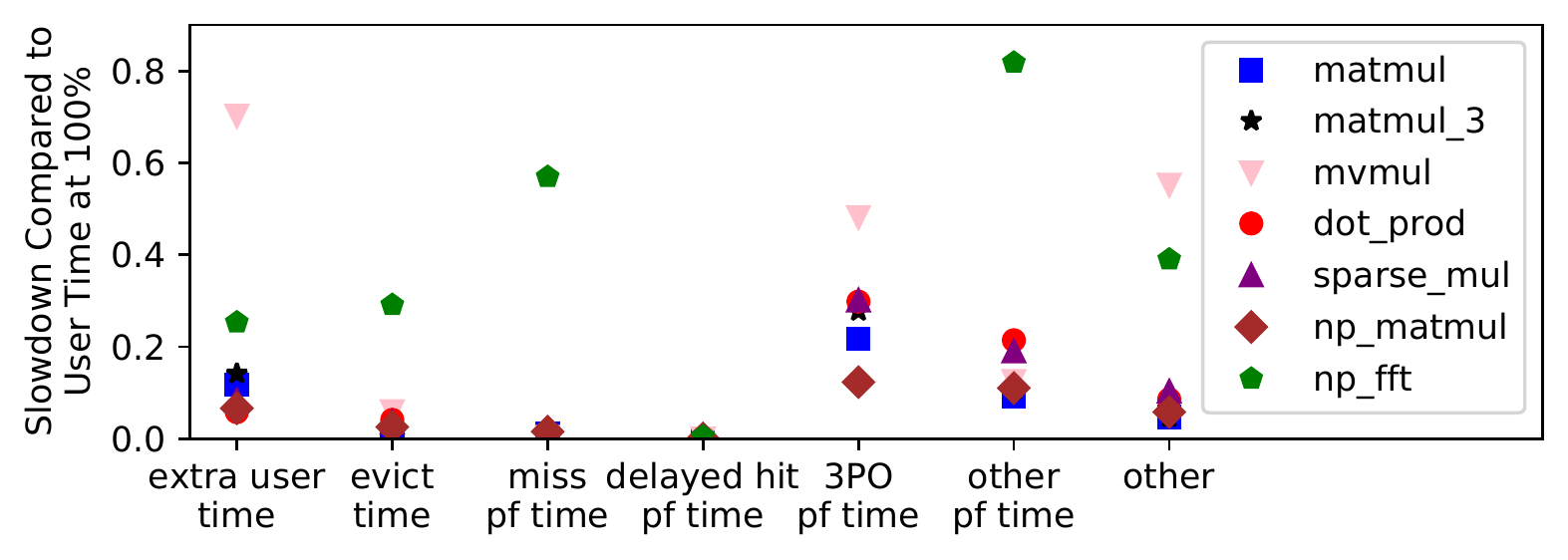}
    \caption{Components of system overhead at 20\% local memory, with \sys{}. \emph{np\_fft}'s ``\sys{} pf time'' is about 2.1; it is cut off in the displayed graph.}
    \label{fig:breakdown_25gb_3po}
\end{figure}

\begin{figure}[t]
    \centering
    \includegraphics[width=\linewidth]{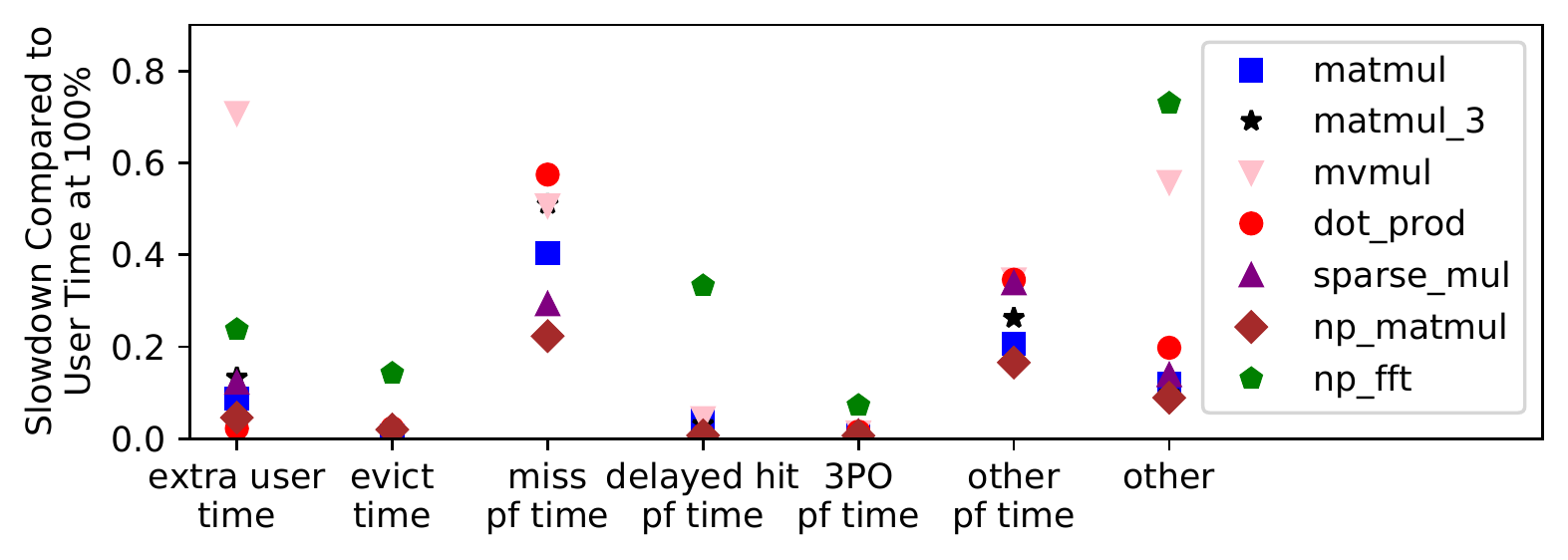}
    \caption{Components of system overhead at 20\% local memory. \emph{np\_fft}'s ``miss pf time'' is about 5.2 and its ``other pf time'' is about 1.8; they are cut off in the displayed graph.}
    \label{fig:breakdown_25gb_linux}
\end{figure}

In previous sections, we saw that performance degrades as the local memory ratio is reduced, even with \sys{}.
This section aims to characterize the sources of this degradation.

\figref{fig:breakdown_25gb_3po} and \figref{fig:breakdown_25gb_linux} show a breakdown of the performance degradation at 20\% local memory on the \emph{25gb} setup into various components, to help determine the relative contribution of each.
Extra user time refers to the amount of additional time spent in userspace when executing at 20\% memory ratio, compared to 100\% memory ratio.
A possible cause is additional cache misses and TLB misses caused by switching to kernel mode to prefetch pages and the high rate of TLB shootdowns~\cite{optimizing_tlb}.
Eviction time is the time that the application is blocked because pages need to be evicted.
Fastswap offloads evictions to a different core, so the application would only block on evictions if the evictions core or network is saturated.
``Miss pf time'' is the time spent on page I/O on a major page fault, and ``delayed hit time'' is the time spent waiting for page I/O to complete in a minor page fault.
\sys{} time is the time in the page fault handler spent on processing for \sys{}'s prefetching.
Finally, ``other pf time'' is the amount of time in the page fault handler not spent waiting for I/O or on processing for \sys{}, and ``other time'' is the remaining time in the system's overall wall clock time (across all cores, for multicore applications) not explained by the previous factors.
We normalize times by the  time spent executing the program in userspace at a 100\% memory ratio.

Eviction time, ``miss pf time'' and ``delayed hit time'' components are small for \sys{}, leading us to conclude that \sys{} can effectively avoid requiring the application to block on I/O.
Therefore, these items do not significantly contribute to \sys{}'s performance degradation at low memory ratios.

One cause of the degradation is overhead that \sys{} does not reduce.
For example, the extra user time is about the same for \sys{} and Linux.
Additionally, a significant amount of degradation, both for \sys{} and for Linux, is due to software overheads in the page fault handler unrelated to \sys{}.
This could be due to reading and updating the state Linux maintains separate from the page table (e.g, swap cache, cgroups, etc.).
Systems that implement prefetching in user space~\cite{aifm, swapping_in_user_space} or via cache coherence~\cite{rethinking} could avoid some of these overheads, though they may suffer from different overheads (e.g., ``software page faults'' cause high overheads from smart pointers~\cite{eleos, aifm}).

Another cause of degradation is overhead of \sys{} itself.
\sys{}'s routines for prefetching and prefaulting pages add overhead not present in Linux, and in some cases, these overheads eat away significantly at the gains from eliminating I/O time.
This is consistent with other observations that software overheads can be significant compared to I/O that completes in microsecond timescales~\cite{killer_us}.

\subsection{CPU and Network Requirements} \label{ss:requirements}
\sys{} enables applications to run at low local memory ratios, but doing so requires extra network bandwidth and CPU usage to handle the high rates of prefetching and evicting. Here we measure the extra CPU cores and network bandwidth required to support low local memory ratios.

Like Fastswap, \sys{} offloads evictions to a separate reclaimer core as much as possible; once the reclaimer is saturated, the application cores must assist with handling evictions.
To measure when  the reclaimer core or the network becomes a bottleneck in \sys{}, we run multiple instances of our applications in parallel and for each memory ratio record the number of application cores that can successfully offload their evictions to the single reclaimer core (or spend less than 5\% of their runtime handling evictions).

\begin{figure}[t]
    \centering
    \includegraphics[width=\linewidth]{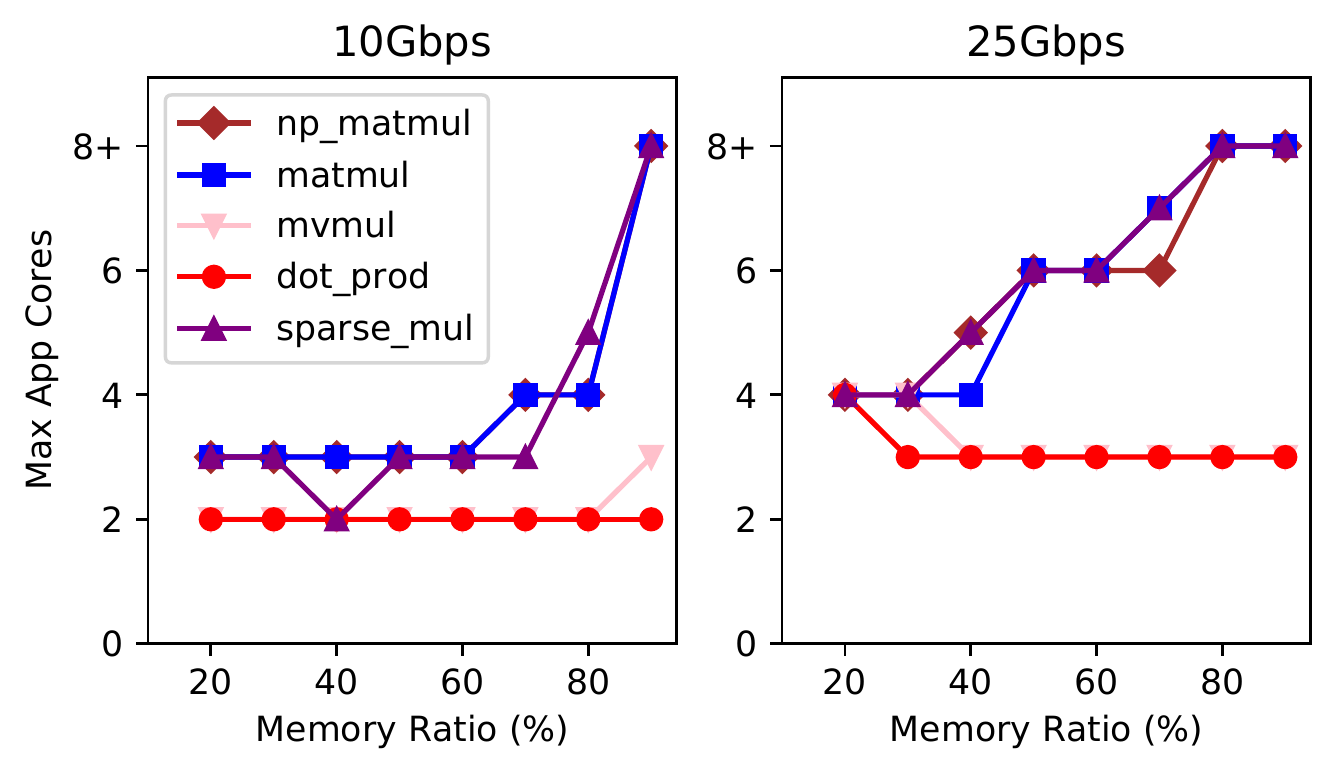}
    \caption{Maximum number of application cores that can be used in parallel before the applications get throttled by handling evictions on their assigned cores.}
    \label{fig:app_cpu_vs_eviction_cpu}
\end{figure}

As shown in~\figref{fig:app_cpu_vs_eviction_cpu}, \sys{} can handle at least 2 and up to 8 or more application cores with a single reclaimer core, depending on the network bandwidth and local memory ratio. The primary bottleneck in this experiment seemed to be the network, rather than the reclaimer core. This is evident from the fact that we can support more application cores with the 25 Gbps network than the 10 Gbps network. In addition, we observed that when application cores start handling evictions, the network bandwidth is close to saturation. However, these results show the worst-case behavior, because all applications start at the same time, aligning their peak bandwidth usage, which is about 50-100\% higher than average usage. Thus, \sys{} could support more applications if their start times were staggered. Because running applications at low memory ratios can impose significant network bandwidth requirements, \sys{} is most suitable for making use of a moderate number of extra cores on a memory-constrained server.

\subsection{Tracing and Post Processing}\label{ss:tape_eval}

\begin{figure}[t]
    \centering
    \includegraphics[width=\linewidth]{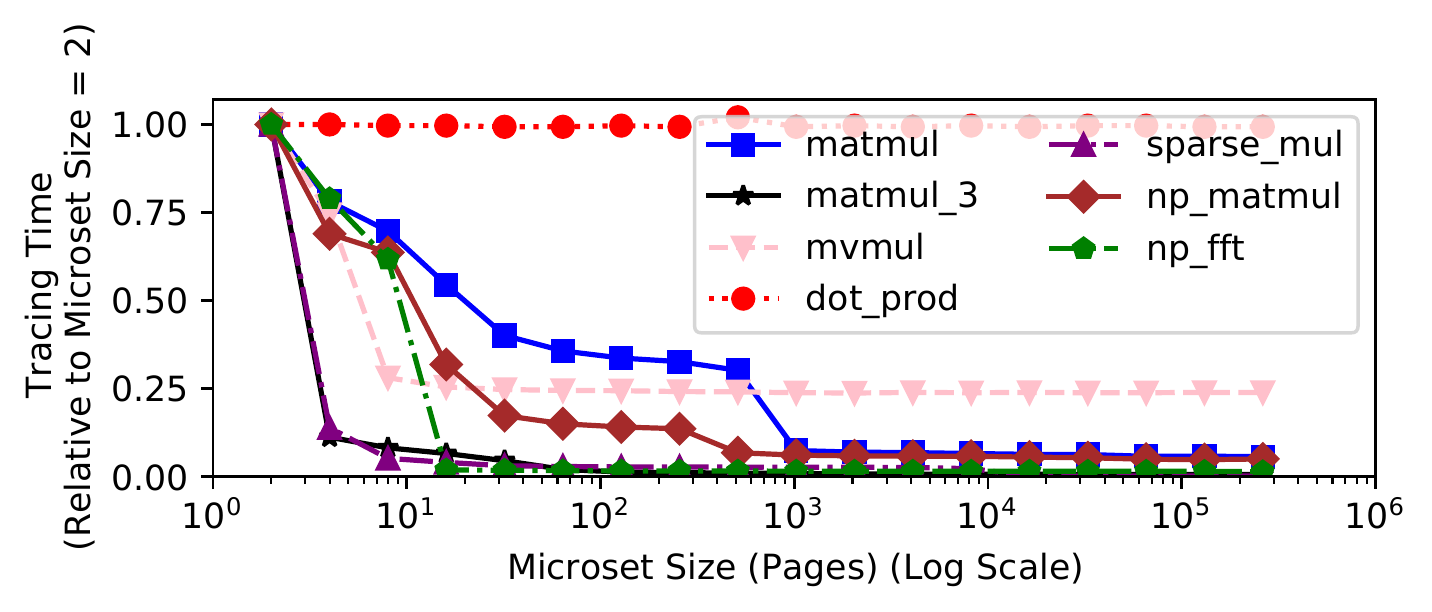}
    \caption{Relative tracing time vs. microset size.}
    \label{fig:tracing_time_vs_microset_size}
\end{figure}

\begin{figure}[t]
    \centering
    \includegraphics[width=\linewidth]{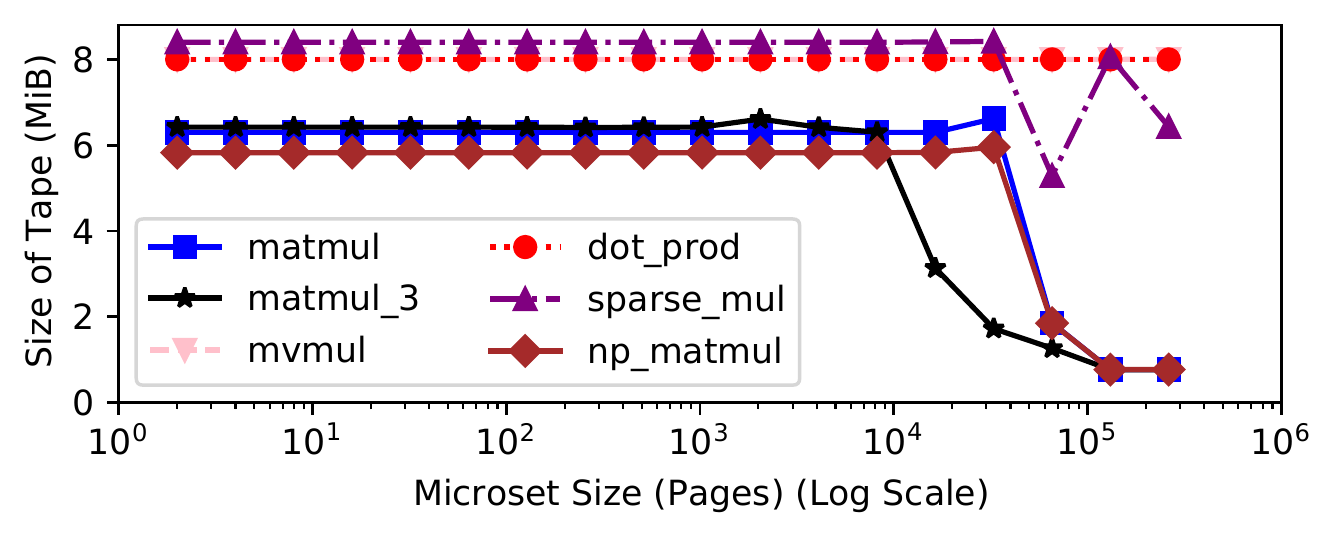}
    \caption{Tape size vs. microset size. \emph{np\_fft} is not shown; it is consistently at about 60 MiB.}
    \label{fig:tape_size_vs_microset_size}
\end{figure}

\begin{figure}[t]
    \centering
    \includegraphics[width=\linewidth]{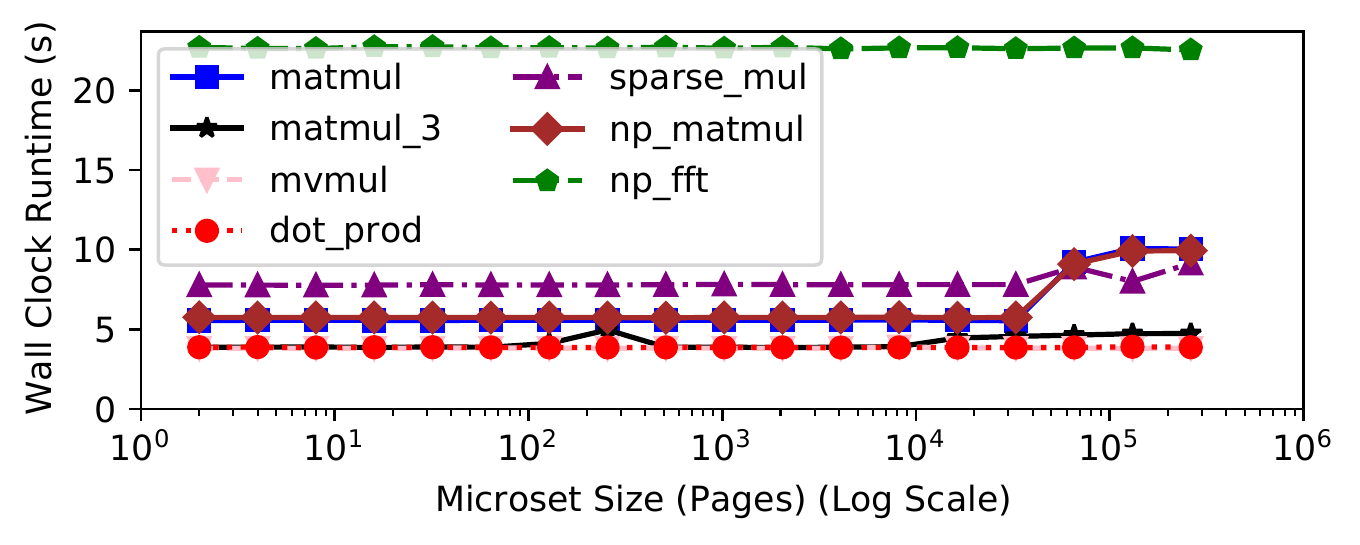}
    \caption{Execution time vs. microset size.}
    \label{fig:runtime_vs_microset_size}
\end{figure}

\begin{table}[t]
    \centering
    \footnotesize
    \begin{tabular}{|c||c|c|c|}\hline
        workload & Tracing Time (s) & Trace Size (MiB) & PostP Time (s)\\\hline\hline
        \textit{matmul} & 4.74 & 17.3 & 2.18\\\hline
        \textit{matmul\_3} & 6.30 & 29.4 & 1.26\\\hline
        \textit{mvmul} & 2.91 & 8.16 & 1.77\\\hline
        \textit{dot\_prod} & 2.94 & 8.00 & 1.77\\\hline
        \textit{sparse\_mul} & 69.2 & 1110 & 102\\\hline
        \textit{np\_matmul} & 4.98 & 16.2 & 2.02\\\hline
        \textit{np\_fft} & 10.5 & 94.9 & 17.1\\\hline
    \end{tabular}
    \vspace{0.2in}
    \caption{Tracing time, trace size, and postprocessing time for microset size 1024, presented to 3 significant figures.}
    \vspace{-.15in}
    \label{tab:stats_uss_1024}
\end{table}

We now aim to understand how \sys{}'s performance is affected by the microset size.
Although the microset size is a parameter of only the tracing phase, the microset size indirectly affects the postprocessing and runtime phases via the size and quality of the produced trace.

The tracing time (\figref{fig:tracing_time_vs_microset_size}) and trace size (not shown) have similar patterns.
Trace size has a similar pattern as tracing time because our tracing algorithm (\secref{ss:microsets}) incurs page faults proportional to the size of the result trace.
Postprocessing time (not shown) has a very similar pattern to trace size because the postprocessing procedure linearly scans the trace, operating on it in a streaming fashion.
These three quantities (tracing time, trace size, and postprocessing time) remain the same or decrease as the microset size is increased, for each workload.
This is because, when the microset size is doubled, the programs' locality causes the number of accesses covered by a microset to increase by \emph{more} than a factor of two.
Programs that do not exhibit locality, like \emph{dot\_prod}, do not become more efficient to trace at larger microset sizes.
Some workloads have one or more ``cliffs'' in the graph, at points where the microsets become large enough to contain larger aspects of the program's structure.
Similar trends have been observed in prior systems (e.g., lifetime curves~\cite{denning1980workingsets}).

The size of the postprocessed tape (\figref{fig:tape_size_vs_microset_size}) follows a different pattern---for most workloads it is flat, with a cliff at high microset sizes.
This is because much of the unnecessary accesses preserved at low microset sizes are ``filtered out'' by the postprocessing.
Thus, the postprocessing allows the tape to be small even at small microset sizes, for which the trace is large.
This suggests that we could rely solely on postprocessing instead of using microsets, but this would be undesirable because larger microset sizes make tracing faster, whereas postprocessing does not.
At very high microset sizes, the tape size decreases.
One explanation for this is that, at these very high microset sizes, the tracing process filters out information that the postprocessing would have preserved.
Indeed, we can see that this ``cliff'' in the tape size often corresponds to an increase in runtime in \figref{fig:runtime_vs_microset_size}.
This may be because the information lost by using these very large microset sizes resulted in lower prefetching quality.

Our goal with microsets in \secref{ss:microsets} was to make it \emph{practical} to trace programs.
Even using a microset size of 2 is borderline impractical for some workloads; for example, for \emph{sparse\_mul}, the trace is $>40$ GiB and takes over 40 minutes to collect.
Using DBI would be much slower than even this.

\begin{figure}[t]
    \centering
    \includegraphics[width=\linewidth]{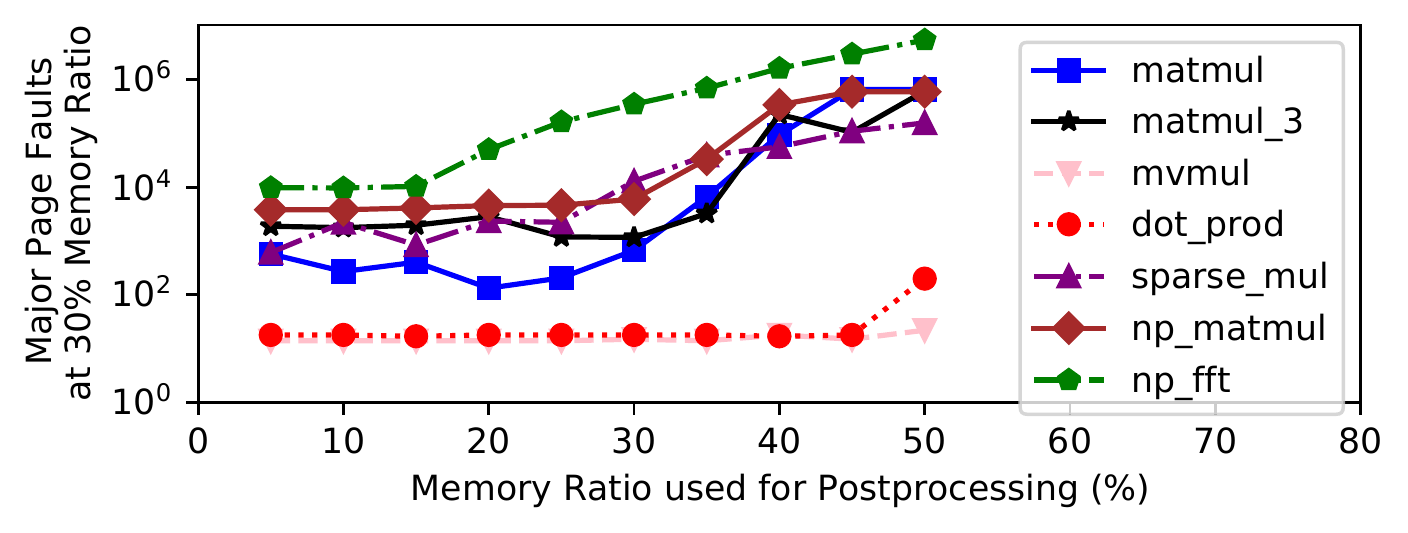}
    \caption{Major page faults (30\% memory ratio) vs. memory ratio used for postprocessing.}
    \label{fig:majflt_vs_postp}
\end{figure}

\sys{} postprocesses the collected trace by simulating an LRU eviction policy, but the actual eviction policy used by Linux at runtime may differ significantly from this (\S\ref{ss:tape}).
To evaluate the impact of using this approximation, we vary the amount of memory used when simulating LRU for postprocessing, while keeping the available memory at runtime fixed.
The results are shown in \figref{fig:majflt_vs_postp}.
For some workloads, like \emph{np\_fft} and \emph{sparse\_mul}, prefetching accuracy increases when postprocessing is done assuming a smaller amount of memory than is actually available at runtime.
This indicates that these workloads are affected by inaccuracies of assuming LRU, which can be somewhat mitigated by assuming less memory when postprocessing to be more conservative.

%% file: 07_related.tex
\section{Related Work}
The prior work most similar to \sys{} is MAGE~\cite{mage}. MAGE computes the memory access pattern of an oblivious program in advance and both prefetches and evicts according to it at runtime.
However, MAGE's techniques for collecting the access pattern and managing memory at runtime are tailored to Secure Computation, a type of cryptographic computation.
For example, MAGE assumes that the program is written in a specific DSL.
In contrast, \sys{} develops techniques that apply to Linux processes that are oblivious and deterministic, but are otherwise generic.

Other prior works use a variety of heuristics for prefetching.
Some perform prefetching by predicting a program's future accesses based on earlier memory accesses, often from within the same execution~\cite{he2013io, soundararajan2008context, diskseen, leap}.
Others perform targeted prefetching based on measurements of a program's working sets from prior executions~\cite{zhang2011fast, reap}.
Some prior works have even used machine learning to predict future memory accesses based on prior ones~\cite{hashemi2018learning, learningrelaxedbelady, lynx}.
Unlike these lines of research, \sys{} prefetches data using the access pattern directly instead of heuristics.

Another line of work aims to allow application developers to include prefetch hints in their program~\cite{vandebogart2009reducing, patterson1995informed, tomkins1997informed}, or otherwise gain increased control of paging from userspace~\cite{extpagecache, aifm}.
For example, the recent AIFM system~\cite{aifm} allows applications to use far memory via ``remoteable'' data structures with custom prefetching policies designed using special AIFM APIs.
In contrast, \sys{} does not require the application developer to extensively modify her code or provide hints.

Another approach to memory management is for the compiler, rather than the application developer, to insert prefetch directives into the program~\cite{mowry1996automatic}.
With this approach, the compiler analyzes the source code to prefetch data for future iterations of a loop, without requiring hints from the application developer.
In contrast, \sys{} operates on Linux processes as a black box.
It is agnostic to the language the program is written in, and can prefetch for the entire application, not just loops.

%% file: 08_conclusion.tex
\section{Conclusion} \label{s:concl}

In this paper we describe \sys{}, a system that enables perfect heuristic-free prefetching from far memory for general oblivious applications. With \sys{}, applications can run at low local memory ratios with significantly less performance degradation than with existing prefetchers such as Linux's prefetcher and state-of-the-art Leap~\cite{leap}. \sys{} also allows us to study the fundamental overheads of paging-based prefetching.

%% file: main.bbl
\begin{thebibliography}{10}

\bibitem{eigen}
Eigen.
\newblock http://eigen.tuxfamily.org.

\bibitem{numpy}
Numpy.
\newblock https://numpy.org/.

\bibitem{remote_regions}
Marcos~K. Aguilera, Nadav Amit, Irina Calciu, Xavier Deguillard, Jayneel
  Gandhi, Stanko Novakovic, Arun Ramanathan, Pratap Subrahmanyam, Lalith
  Suresh, Kiran Tati, Rajesh Venkatasubramanian, and Michael Wei.
\newblock Remote regions: A simple abstraction for remote memory.
\newblock USENIX ATC '18, pages 775--787, 2018.

\bibitem{leap}
Hasan Al~Maruf and Mosharaf Chowdhury.
\newblock Effectively prefetching remote memory with {Leap}.
\newblock In {\em 2020 $\{$USENIX$\}$ Annual Technical Conference
  ($\{$USENIX$\}$$\{$ATC$\}$ 20)}, pages 843--857, 2020.

\bibitem{cfm}
Emmanuel Amaro, Christopher Branner-Augmon, Zhihong Luo, Amy Ousterhout,
  Marcos~K Aguilera, Aurojit Panda, Sylvia Ratnasamy, and Scott Shenker.
\newblock Can far memory improve job throughput?
\newblock In {\em Proceedings of the Fifteenth European Conference on Computer
  Systems}, pages 1--16, 2020.

\bibitem{optimizing_tlb}
Nadav Amit.
\newblock Optimizing the $\{$TLB$\}$ shootdown algorithm with page access
  tracking.
\newblock In {\em 2017 $\{$USENIX$\}$ Annual Technical Conference
  ($\{$USENIX$\}$$\{$ATC$\}$ 17)}, pages 27--39, 2017.

\bibitem{memorybandwidth}
Anandtech.
\newblock Amd 3rd gen epyc milan review: A peak vs per core performance
  balance.
\newblock Accessed: October 9, 2021.

\bibitem{fastesthdd}
Arstechnica.
\newblock Seagate’s new mach.2 is the world’s fastest conventional hard
  drive.
\newblock Accessed: October 9, 2021.

\bibitem{delayed_hits}
Nirav Atre, Justine Sherry, Weina Wang, and Daniel~S Berger.
\newblock Caching with delayed hits.
\newblock In {\em Proceedings of the Annual conference of the ACM Special
  Interest Group on Data Communication on the applications, technologies,
  architectures, and protocols for computer communication}, pages 495--513,
  2020.

\bibitem{killer_us}
Luiz Barroso, Mike Marty, David Patterson, and Parthasarathy Ranganathan.
\newblock Attack of the killer microseconds.
\newblock {\em Communications of the ACM}, 60(4):48--54, 2017.

\bibitem{datacenter_as_a_computer}
Luiz~Andr{\'e} Barroso and Urs H{\"o}lzle.
\newblock The datacenter as a computer: An introduction to the design of
  warehouse-scale machines.
\newblock {\em Synthesis lectures on computer architecture}, 4(1):1--108, 2009.

\bibitem{beladys}
Laszlo~A. Belady.
\newblock A study of replacement algorithms for a virtual-storage computer.
\newblock {\em IBM Systems journal}, 5(2):78--101, 1966.

\bibitem{linux_kernel}
Daniel~Pierre Bovet and Marco Cesati.
\newblock {Page frame reclaiming}.
\newblock In {\em {Understanding the Linux Kernel}}, chapter~17, page 679.
  O’Reilly Media, 2006.

\bibitem{transparent_dynamic}
Derek Bruening, Qin Zhao, and Saman Amarasinghe.
\newblock Transparent dynamic instrumentation.
\newblock In {\em Proceedings of the 8th ACM SIGPLAN/SIGOPS conference on
  Virtual Execution Environments}, pages 133--144, 2012.

\bibitem{rethinking}
Irina Calciu, M~Talha Imran, Ivan Puddu, Sanidhya Kashyap, Hasan~Al Maruf, Onur
  Mutlu, and Aasheesh Kolli.
\newblock Rethinking software runtimes for disaggregated memory.
\newblock In {\em Proceedings of the 26th ACM International Conference on
  Architectural Support for Programming Languages and Operating Systems}, pages
  79--92, 2021.

\bibitem{denning1980workingsets}
P.~J. Denning.
\newblock Working sets past and present.
\newblock {\em IEEE Trans. Softw. Eng.}, SE-6(1), 1980.

\bibitem{diskseen}
Xiaoning Ding, Song Jiang, Feng Chen, Kei Davis, and Xiaodong Zhang.
\newblock Diskseen: Exploiting disk layout and access history to enhance i/o
  prefetch.
\newblock In {\em USENIX Annual Technical Conference}, volume~7, pages
  261--274, 2007.

\bibitem{oram}
O.~Goldreich and R.~Ostrovsky.
\newblock Software protection and simulation on oblivious {RAM}s.
\newblock {\em J. ACM}, 43(3), 1996.

\bibitem{infiniswap}
J.~Gu, Y.~Lee, Y.~Zhang, M.~Chowdhury, and K.~G. Shin.
\newblock Efficient memory disaggregation with infiniswap.
\newblock In {\em NSDI}. USENIX, 2017.

\bibitem{extpagecache}
K.~Harty and D.~R. Cheriton.
\newblock Application-controlled physical memory using external page-cache
  management.
\newblock In {\em ASPLOS}. ACM, 1992.

\bibitem{hashemi2018learning}
M.~Hashemi, K.~Swersky, J.~A. Smith, G.~Ayers, H.~Litz, J.~Chang, C.~Kozyrakis,
  and P.~Ranganathan.
\newblock Learning memory access patterns.
\newblock In {\em ICML}, 2018.

\bibitem{he2013io}
J.~He, J.~Bent, A.~Torres, G.~Grider, G.~Gibson, C.~Maltzahn, and X.-H. Sun.
\newblock {I/O} acceleration with pattern detection.
\newblock In {\em HPDC}. ACM, 2015.

\bibitem{optaneperf}
Joseph Izraelevitz, Jian Yang, Lu~Zhang, Juno Kim, Xiao Liu, Amirsaman
  Memaripour, Yun~Joon Soh, Zixuan Wang, Yi~Xu, Subramanya~R Dulloor, et~al.
\newblock Basic performance measurements of the intel optane dc persistent
  memory module.
\newblock {\em arXiv preprint arXiv:1903.05714}, 2019.

\bibitem{kang2014co}
Uksong Kang, Hak-Soo Yu, Churoo Park, Hongzhong Zheng, John Halbert, Kuljit
  Bains, S.~Jang, and Joo~Sun Choi.
\newblock Co-architecting controllers and dram to enhance dram process scaling.
\newblock In {\em The memory forum}, 2014.

\bibitem{mage}
Sam Kumar, David~E. Culler, and Raluca~Ada Popa.
\newblock {MAGE}: Nearly zero-cost virtual memory for secure computation.
\newblock In {\em 15th $\{$USENIX$\}$ Symposium on Operating Systems Design and
  Implementation ($\{$OSDI$\}$ 21)}, 2021.

\bibitem{lynx}
A.~Laga, J~Boukhobza, M.~Koskas, and F.~Singhoff.
\newblock {Lynx}: A learning {Linux} prefetching mechanism for {SSD}
  performance model.
\newblock In {\em NVMSA}. IEEE, 2016.

\bibitem{softwaredefinedfarmem}
Andres Lagar-Cavilla, Junwhan Ahn, Suleiman Souhlal, Neha Agarwal, Radoslaw
  Burny, Shakeel Butt, Jichuan Chang, Ashwin Chaugule, Nan Deng, Junaid Shahid,
  et~al.
\newblock Software-defined far memory in warehouse-scale computers.
\newblock In {\em Proceedings of the Twenty-Fourth International Conference on
  Architectural Support for Programming Languages and Operating Systems}, pages
  317--330, 2019.

\bibitem{java_fork_join}
Doug Lea.
\newblock A java fork/join framework.
\newblock In {\em Proceedings of the ACM 2000 conference on Java Grande}, pages
  36--43, 2000.

\bibitem{lee2016technology}
Seok-Hee Lee.
\newblock Technology scaling challenges and opportunities of memory devices.
\newblock In {\em IEEE International Electron Devices Meeting}, 2016.

\bibitem{liang2005swapping}
Shuang Liang, Ranjit Noronha, and Dhabaleswar~K Panda.
\newblock Swapping to remote memory over infiniband: An approach using a high
  performance network block device.
\newblock In {\em 2005 IEEE International Conference on Cluster Computing},
  pages 1--10. IEEE, 2005.

\bibitem{traceoblivious}
C.~Liu, M.~Hicks, and E.~Shi.
\newblock Memory trace oblivious program execution.
\newblock In {\em Computer Security Foundations Symposium}. IEEE, 2013.

\bibitem{pin}
C.-K. Luk, R.~Cohn, R.~Muth, H.~Patil, A.~Klauser, G.~Lowney, S.~Wallace, V.~J.
  Reddi, and K.~Hazelwood.
\newblock {Pin}: Building customized program analysis tools with dynamic
  instrumentation.
\newblock In {\em PLDI}. ACM, 2005.

\bibitem{mowry1996automatic}
T.~C. Mowry, A.~K. Demke, and O.~Krieger.
\newblock Automatic compiler-inserted {I/O} prefetching for out-of-core
  applications.
\newblock In {\em OSDI}. USENIX, 1996.

\bibitem{valgrind}
N.~Nethercote and J.~Seward.
\newblock Valgrind: A framework for heavyweight dynamic binary instrumentation.
\newblock In {\em PLDI}. ACM, 2007.

\bibitem{eleos}
Meni Orenbach, Pavel Lifshits, Marina Minkin, and Mark Silberstein.
\newblock Eleos: Exitless os services for sgx enclaves.
\newblock In {\em Proceedings of the Twelfth European Conference on Computer
  Systems}, pages 238--253, 2017.

\bibitem{patterson1995informed}
R.~H. Patterson, G.~A. Gibson, E.~Ginting, D.~Stodolsky, and J.~Zelenka.
\newblock Informed caching and prefetching.
\newblock In {\em SOSP}. ACM, 1995.

\bibitem{informed_prefetching}
R~Hugo Patterson, Garth~A Gibson, Eka Ginting, Daniel Stodolsky, and Jim
  Zelenka.
\newblock Informed prefetching and caching.
\newblock In {\em Proceedings of the fifteenth ACM symposium on Operating
  systems principles}, pages 79--95, 1995.

\bibitem{intel_tbb}
James Reinders.
\newblock {\em {Intel Threading Building Blocks: Outfitting C++ for Multi-Core
  Processor Parallelism}}.
\newblock 2007.

\bibitem{aifm}
Zhenyuan Ruan, Malte Schwarzkopf, Marcos~K Aguilera, and Adam Belay.
\newblock $\{$AIFM$\}$: High-performance, application-integrated far memory.
\newblock In {\em 14th $\{$USENIX$\}$ Symposium on Operating Systems Design and
  Implementation ($\{$OSDI$\}$ 20)}, pages 315--332, 2020.

\bibitem{legoos}
Yizhou Shan, Yutong Huang, Yilun Chen, and Yiying Zhang.
\newblock Legoos: A disseminated, distributed os for hardware resource
  disaggregation.
\newblock OSDI'18, pages 69--87, 2018.

\bibitem{learningrelaxedbelady}
Z.~Song, D.~S. Berger, K.~Li, and W.~Lloyd.
\newblock Learning relaxed {Belady} for content distribution network caching.
\newblock In {\em NSDI}. USENIX, 2020.

\bibitem{soundararajan2008context}
G.~Soundararajan, M.~Mihailescu, and C.~Amza.
\newblock Context-aware prefetching at the storage server.
\newblock In {\em ATC}. USENIX, 2008.

\bibitem{go}
{The Go Community}.
\newblock The go programming language.
\newblock https://golang.org.

\bibitem{openmp}
{The Go Community}.
\newblock Openmp.
\newblock https://www.openmp.org/.

\bibitem{tomkins1997informed}
A.~Tomkins, R.~H. Patterson, and G.~Gibson.
\newblock Informed multi-process prefetching and caching.
\newblock In {\em Sigmetrics}. ACM, 1997.

\bibitem{informed_multi_process}
Andrew Tomkins, R~Hugo Patterson, and Garth Gibson.
\newblock Informed multi-process prefetching and caching.
\newblock {\em ACM SIGMETRICS Performance Evaluation Review}, 25(1):100--114,
  1997.

\bibitem{reap}
D.~Ustiugov, P.~Petrov, M.~Kogias, E.~Bugnion, and B.~Grot.
\newblock Benchmarking, analysis, and optimization of serverless function
  snapshots.
\newblock In {\em ASPLOS}. ACM, 2021.

\bibitem{vandebogart2009reducing}
S.~VanDeBogart, C.~Frost, and E.~Kohler.
\newblock Reducing seek overhead with application-directed prefetching.
\newblock In {\em ATC}. USENIX, 2009.

\bibitem{app_directed}
Steve VanDeBogart, Christopher Frost, and Eddie Kohler.
\newblock Reducing seek overhead with application-directed prefetching.
\newblock In {\em USENIX Annual Technical Conference}, 2009.

\bibitem{obliviousds}
X.~S. Wang, K.~Nayak, C.~Liu, T.-H.~H. Chan, E.~Shi, E.~Stefanov, and Y.~Huang.
\newblock Oblivious data structures.
\newblock In {\em CCS}. ACM, 2014.

\bibitem{spark}
Matei Zaharia, Mosharaf Chowdhury, Tathagata Das, Ankur Dave, Justin Ma, Murphy
  McCauly, Michael~J Franklin, Scott Shenker, and Ion Stoica.
\newblock Resilient distributed datasets: A fault-tolerant abstraction for
  in-memory cluster computing.
\newblock In {\em 9th $\{$USENIX$\}$ Symposium on Networked Systems Design and
  Implementation ($\{$NSDI$\}$ 12)}, pages 15--28, 2012.

\bibitem{zhang2011fast}
I.~Zhang, A.~Garthwaite, Y.~Baskakov, and K.~C. Barr.
\newblock Fast restore of checkpointed memory using working set estimation.
\newblock In {\em VEE}. ACM, 2011.

\bibitem{opaque}
Wenting Zheng, Ankur Dave, Jethro~G Beekman, Raluca~Ada Popa, Joseph~E
  Gonzalez, and Ion Stoica.
\newblock Opaque: An oblivious and encrypted distributed analytics platform.
\newblock In {\em 14th $\{$USENIX$\}$ Symposium on Networked Systems Design and
  Implementation ($\{$NSDI$\}$ 17)}, pages 283--298, 2017.

\bibitem{swapping_in_user_space}
Kan Zhong, Wenlin Cui, Youyou Lu, Quanzhang Liu, Xiaodan Yan, Qizhao Yuan,
  Siwei Luo, and Keji Huang.
\newblock Revisiting swapping in user-space with lightweight threading.
\newblock {\em arXiv preprint arXiv:2107.13848}, 2021.

\end{thebibliography}
